\documentclass[a4paper,fleqn]{cas-sc}

\usepackage[normalem]{ulem}
\usepackage[dvipsnames]{xcolor}
\usepackage{graphicx}
\usepackage{amsmath}
\usepackage{amsfonts}
\usepackage{amscd}
\usepackage{amssymb}
\usepackage{zref-clever}
\zcsetup{cap,abbrev}
\usepackage[colorlinks]{hyperref}
\usepackage{braket}
\usepackage{enumitem}
\usepackage{multirow}
\usepackage{bm}
\usepackage{bbm}
\usepackage{units}
\usepackage{grffile}
\usepackage{mathtools}
\usepackage{xstring}
\usepackage{xspace}
\usepackage{afterpage}

\numberwithin{equation}{section}

\DeclareMathOperator\sech{sech}

\DeclareMathOperator\tr{tr}
\DeclareMathOperator{\range}{range}

\newcommand{\cref}[1]{\zcref{#1}}
\newcommand{\Cref}[1]{\zcref[S]{#1}}

\newcommand{\nmax}{{n_\text{max}}}
\newcommand{\ns}[0]{s}
\newcommand{\nl}[0]{\ell}

\newcommand{\nn}[1]{\hat n^{#1}}

\newcommand{\Tnew}{T}
\newcommand{\Ttrunc}{\bar T}
\newcommand\jtrunc{\bar j} 
\newcommand\fatj{j} 

\newcommand\vconfiguration{valid configuration\xspace}

\newcommand{\SU}{\text{SU}}

\newcommand{\Order}[1]{\mathcal{O}(#1)}

\newcommand\aux\chi
\newcommand\pif{\int\limits_\chi}
\newcommand\Nf{N_\text{f}}
\newcommand\Nc{N_\text{c}}

\newcommand\SG{S^\text{G}}

\newcommand{\f}{F}
\newcommand\mub{\mu_\text{B}}
\newcommand\mui{\mu_\text{I}}

\newcommand{\X}{X}
\newcommand{\Y}{Y}
\newcommand{\nR}{n_\text{R}}

\newcommand\Vs{V_\text{s}}
\newcommand\muc{\mu^{\text{c}}}
\newcommand\rhob{\rho_\text{B}}

\renewcommand\Zbar{\bar Z}
\renewcommand\Lt{L_\text{t}}
\renewcommand{\epsilon}{\varepsilon}
\renewcommand\j{j}

\begin{document}

\small

\title[mode=title]{Order-separated tensor-network method for QCD in the strong-coupling expansion}
\author[1]{Thomas Samberger}[orcid=0009-0002-0445-386X]
\ead{thomas.samberger@ur.de}
\author[1]{Jacques Bloch}[orcid=0000-0002-8443-4804]
\ead{jacques.bloch@ur.de}
\author{Robert Lohmayer}[orcid=0000-0001-6207-4695]
\ead{robert.lohmayer@ur.de}
\author[1]{Tilo Wettig}[orcid=0000-0001-6732-9204]
\ead{tilo.wettig@ur.de}

\shorttitle{Order-separated tensor-network method for QCD in the strong-coupling expansion}
\shortauthors{T. Samberger, J. Bloch, R. Lohmayer, T. Wettig}

\address[1]{Institute for Theoretical Physics, University of Regensburg, 93040 Regensburg, Germany}

\date{\today}

\begin{abstract}
We introduce the order-separated Grassmann higher-order tensor renormalization group (OS-GHOTRG) method for QCD with staggered quarks in the strong-coupling expansion. The method allows us to determine the expansion coefficients of the partition function, from which we can obtain the strong-coupling expansions of thermodynamical observables. 
We use the method in two dimensions to compute the free energy, the particle-number density, and the chiral condensate  as a function of the chemical potential up to third order in the inverse coupling $\beta$. Although near the phase transition the expansion is only a good approximation to the full theory at small $\beta$, we show that the range of applicability can be greatly extended by fits to judiciously chosen transition functions.
\end{abstract}

\maketitle

\tableofcontents

\allowdisplaybreaks[4]

\section{Introduction}

This paper is a sequel to our earlier paper \cite{Samberger:2025hsr}. Our long-term goal is to develop a numerical method that can be used to compute the phase diagram of Quantum Chromodynamics (QCD). Lattice QCD is mainly investigated using Monte Carlo (MC) sampling. However, in the presence of a quark chemical potential $\mu$, the action becomes complex and the weights in the partition function are no longer positive, so that the standard MC method can no longer be used. A number of methods were developed to circumvent the sign problem, e.g., reweighted MC, complex Langevin, thimbles and path optimization, density of states, Taylor expansions around $\mu=0$, continuation from imaginary chemical potential, and others, see \cite{deForcrand:2009zkb,Aarts:2015tyj} for reviews. Unfortunately,  all of these methods fail when the ratio of chemical potential to temperature becomes larger than one, and thus only a small region of the phase diagram is accessible. The typical signal for the failure of these methods is that the sign problem grows, and as a result the method becomes exponentially expensive in the volume or breaks down completely. An approach of a somewhat different kind is that of dual variables, where the original degrees of freedom in the partition function are integrated out in favor of new variables that are often called occupation numbers. The advantage of this reformulation is that the sign problem typically becomes much milder, which is generally understood to be due to the fact that the partition function is now expressed in terms of the physical degrees of freedom at low energy, i.e., baryon loops as well as mesonic and mass contributions. The dual formulation was investigated in detail in the infinite-coupling limit with MC methods \cite{Rossi:1984cv,Karsch:1988zx,Fromm:2010lga,deForcrand:2009dh},  where the sampling of loop configurations is typically performed using the worm algorithm \cite{Prokofiev:2001zz}. First attempts to go beyond the infinite-coupling limit in the dual formulation were made using the worm algorithm \cite{Forcrand2014} and vertex models \cite{Gagliardi:2019cpa,Kim:2023dnq}. 

In the dual formulation, the partition function can often be rewritten as the complete trace of a tensor network, with local tensors distributed over the sites of the lattice. Tensor-network methods were developed in recent years to compute such partition functions and related observables. 
These methods could play an important role in the investigation of theories with complex actions, as they could potentially overcome the sign problem, e.g., in  simulations of lattice QCD at $\mu\ne0$. There are two main reasons for this. First, the tensor formulation describes the theory in dual variables, which is known to drastically reduce the sign problem as mentioned above. Second, any remaining sign problem can, in principle, be handled by the tensor-network formulation as the latter is not based on a stochastic sampling of the states in the partition function, but rather makes algebraic approximations to the large tensors encountered in the computation of the partition function.

To compute partition functions using tensor networks, various methods have been developed in recent years. The tensor renormalization group (TRG) method, which is based on singular value decomposition (SVD), was proposed by Levin and Nave \cite{Levin:2006jai} to treat two-dimensional models. The higher-order tensor renormalization group (HOTRG) method, which is based on higher-order singular value decompositions (HOSVD) \cite{DeLathauwer2000}, is an alternative method developed by Xie et al. \cite{Xie_2012}, which can also be used in higher dimensions. The general idea of these methods is to iteratively coarsen the lattice by contracting pairs of adjacent tensors, until only a single tensor is left. Its trace is then the partition function to be computed. During the blocking procedure, the bond dimension $D$ of the coarse-grid tensor, i.e., the range of the tensor indices, rapidly increases. To avoid the curse of dimensionality, the increased bond dimension is reduced using SVD or HOSVD after each contraction step. These methods have a cost that is logarithmic in the volume and polynomial in the bond dimension. The latter is an adjustable parameter that controls the quality of the approximation and crucially affects the accuracy of the results. 
Thermodynamical observables can then be computed by methods based on finite differences or by impurity methods, which involve analytical derivatives of the local tensor. The tensor-network methods just described were successfully applied to various spin and gauge models, e.g., the Ising model in two, three, and four dimensions, the O(2) model, the Z$_3$ model, and others, see \cite{Meurice:2020pxc} for a review. Even systems with a complex action were successfully studied, e.g., the three-dimensional O(2) model with a chemical potential \cite{Bloch:2021mjw}.

In theories with Grassmann variables, the latter can sometimes be integrated out completely during the construction of the tensor network, as for example in the U($N$) gauge theory at infinite coupling \cite{Rossi:1984cv,Milde:2021vln}, in which case the usual TRG or HOTRG method can be used. However, in general the integration of Grassmann variables produces nonlocal sign factors which cannot be included in a tensor-network formulation of the partition function. This can, for example, be observed in the meson-baryon-loop representation of the partition function in QCD at infinite coupling \cite{Rossi:1984cv,Karsch:1988zx}. In such cases the tensor-network formalism is extended by writing the local tensors as products of a numerical and a Grassmann tensor \cite{Gu:2010yh}. During the iterative coarsening procedure the contraction of Grassmann tensors on adjacent sites is performed analytically, giving rise to local sign factors that are absorbed in the numerical tensors, while the numerical tensors are treated as before, i.e., they are contracted and truncated using SVD or HOSVD. These principles were applied to the TRG and HOTRG methods, leading to their Grassmann counter-parts called Grassmann TRG (GTRG) \cite{Shimizu:2014uva,Takeda:2014vwa} and Grassmann HOTRG (GHOTRG) \cite{Sakai:2017jwp}.

In the context of QCD, our group implemented the GHOTRG ideas to investigate QCD at infinite coupling in two dimensions with staggered quarks at nonzero chemical potential \cite{Bloch:2022vqz} and also in three and four dimensions \cite{Bloch:2022yiq,Milde2023}. The method was used to compute the chiral condensate as a function of quark mass and volume, and to confirm the absence of dynamical chiral symmetry breaking in the two-dimensional case. Tensor methods are especially well-suited for this investigation, since they can be used for the very large volumes that are required for small masses.
Furthermore, the quark-number density was computed as a function of the chemical potential, which hinted at a first-order phase transition. 
This study was extended to four dimensions to investigate the phase diagram of infinite-coupling QCD, with quite promising results \cite{Milde2023}. Similar computations were performed very recently to investigate cold and dense QCD in the infinite-coupling limit \cite{Sugimoto:2026wnw}. 

The natural next step is to go beyond the infinite-coupling limit. First attempts to include the full gauge action of two-color QCD in two dimensions were made using numerical integrations of the gauge fields \cite{Asaduzzaman:2023pyz,Pai:2024tip}.
As an alternative approach, we have recently derived a tensor-network formulation for the partition function of lattice QCD in the strong-coupling expansion \cite{Samberger:2025hsr,Samberger2026}. The formulation is valid to any order in the inverse coupling $\beta$, number of dimensions, number of colors, and number of staggered fermion flavors. 
If we truncate the initial local tensor at some order $\beta^\nmax$ we can use the GHOTRG method for QCD \cite{Bloch:2022vqz} without further ado to compute the partition function and thermodynamical observables.\footnote{In contrast to the infinite-coupling case, the new colorless auxiliary Grassmann variables introduced in the GHOTRG method no longer correspond to baryonic degrees of freedom, but are generic fermionic degrees of freedom. Nevertheless, the GHOTRG blocking procedure remains completely identical to that of the infinite-coupling case detailed in \cite{Bloch:2022vqz}.} However, as was shown in \cite{Samberger:2025hsr}, the strong-coupling expansion results obtained with the GHOTRG method quickly break down as $\beta$ is increased because the tensor contractions generate contributions of order higher than $\beta^\nmax$, which are incomplete and were shown to lead to unphysical results.
To remedy this problem we developed the order-separated GHOTRG (OS-GHOTRG) method to compute the coefficients of the strong-coupling expansion of the partition function order by order up to $\beta^\nmax$. The main idea is that each tensor entry is written as a power series in $\beta$ during the iterative procedure. This allows us to remove the unwanted contributions of higher order. Using the expansion of the partition function we can also expand the free energy, i.e., the logarithm of the partition function, and hence the thermodynamical observables, order by order also up to $\beta^\nmax$. 
If the initial local tensor is exact to order $\beta^\nmax$, then the tensor-network results are also exact to that order, up to the truncation errors introduced by the HOSVD approximations.
As was shown in \cite{Samberger:2025hsr}, these expansions yield much better agreement with benchmark data, e.g., obtained from MC simulations. In the present paper we give a detailed description of the OS-GHOTRG method (see also \cite{Samberger:2025dmp} for an early account).

This paper is structured as follows. In \cref{sec:TN} we present the tensor network derived in \cite{Samberger:2025hsr}. In \cref{Sec:Separation} we explain the OS-GHOTRG method in detail. In \cref{sec:trans_inv} we use translational invariance in the partition function to improve the efficiency of the OS-GHOTRG method. In \cref{sec:results} we present a number of results obtained for \SU(3) in two dimensions, up to $\nmax=3$ for one flavor and $\nmax=1$ for two flavors. We validate the tensor-network method by comparing with analytical results for a $2\times2$ lattice and with MC results for an $8\times8$ lattice. We also discuss the applicability of the strong-coupling expansion. For larger lattices we introduce a $\tanh$ model which describes the data very well close to the phase transition and allows for an extrapolation to larger $\beta$ values. In \cref{sec:conclusions} we conclude and give an outlook to future work. In the appendices we provide a number of technical details.

\section{Tensor network for QCD in the strong-coupling expansion}
\label{sec:TN}

\subsection{Partition function}
\label{sec:partition_function}

\begin{figure}
  \centering
  \includegraphics{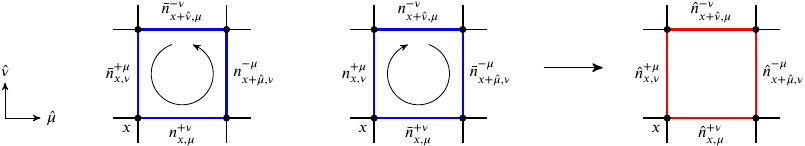}
  \caption{Blue: Edge variables $n$ and $\bar n$ for the oriented plaquettes starting at site $x$ in the $\mu\nu$ and $\nu\mu$ directions, respectively.
    The subscript indicates the link, while the superscript indicates the direction (including sign) in which the plaquette extends perpendicular to the link. For a \vconfiguration the four edge variables around an oriented plaquette must be equal.
    Red: Edge occupations of the unoriented plaquette as defined in \eqref{eq:def_plaq_excitation}, which must also be equal for a \vconfiguration.}
  \label{fig:plaq_to_link}
\end{figure}

In our previous paper \cite{Samberger:2025hsr} we constructed a tensor-network representation for the partition function of the strong-coupling expansion of an $\SU(\Nc)$ gauge theory with $\Nf$ flavors of staggered fermions with masses $m_f$ and quark chemical potentials $\mu_f$ ($f=1,\ldots,\Nf$) on a $d$-dimensional Euclidean space-time lattice.
The major steps in the derivation were: Taylor expansions of all exponentials involving the Wilson plaquette action, the fermion action, and the mass term, integration of all gauge links and all quark Grassmann variables, and finally contraction of all color indices. In the process, one new colorless auxiliary Grassmann variable per link was introduced, independently of the number of colors and flavors. 
The partition function could then be written as the complete trace of a tensor network consisting of local tensors containing a numerical and a Grassmann part,
\begin{equation}\label{eq:final_tensor_network}
  Z = \sum_{\bm{j}}\pif \prod_x (T_x)_{j_{x,-1}j_{x,1}\dots j_{x,-d}j_{x,d}} (K_x)_{F_{x,-1}F_{x,1}\dots F_{x,-d}F_{x,d}} \,,
\end{equation}
where $x$ enumerates the sites of the lattice with lattice volume $V$, the different directions are denoted by $\mu\in\{1,\ldots,d\}$, and $\bm{j}\equiv(j_{x,\mu}| \,\forall\, x, \mu)$ contains all compound link indices $j_{x,\mu}$, which are defined in \cite[Sec.~6.2]{Samberger:2025hsr}. The unoriented link between $x$ and $x+\hat\mu$ is denoted by $(x,\mu)$, and the link identity $(x,-\mu)=(x-\hat\mu,\mu)$ holds.\footnote{Throughout the paper we us the short-hand notation $f_{x,\mu}\equiv f_{(x,\mu)}$ for quantities defined on the link $(x,\mu)$, of which $j_{x,\mu}$ is an example.}
The compound link index $j_{x,\mu}$ contains, among others, the so-called edge variables, see \cref{fig:plaq_to_link} (left), which are nonnegative integers. 
In the following we will use the term ``\vconfiguration'' for a configuration $\bm{j}$ that makes a nonzero contribution to the partition function.

The local numerical tensor $T_x$ depends on the index tuple $\j_x\equiv(j_{x,-1},j_{x,1},\dots, j_{x,-d},j_{x,d})$ and is given in Ref.~\cite[Eq.~(6.16)]{Samberger:2025hsr}.  
We do not give its explicit expression here as it involves many details that are not relevant for the present discussion. 
Its only property that is significant for the derivation of the OS-GHOTRG method is the $\beta$ dependence of its entries, which comes from the Taylor expansion of the exponential of the plaquette action and is given by
\begin{equation}\label{Tbetadep}
(T_x)_{\j_x} \propto
\beta^{n_x}\quad\text{with}\quad n_x(j_x)\equiv\frac18\sum_{{\mu,\nu=\pm1}\atop{|\mu|\neq|\nu|}}^{\pm d} \nn{\nu}_{x,\mu}\,,
\end{equation}
where we have defined the edge-occupation numbers (or edge occupations in short) 
\begin{equation}\label{eq:def_plaq_excitation}
  \nn{\nu}_{x,\mu}(j_{x,\mu})\equiv n^{\nu}_{x,\mu}+\bar n^{\nu}_{x,\mu}
\end{equation}
as the sum of edge variables of two superimposed plaquettes with opposite orientations, see \cref{fig:plaq_to_link} (right).
The edge occupations therefore refer to unoriented plaquettes. 

\begin{figure}
  \centering
  \includegraphics{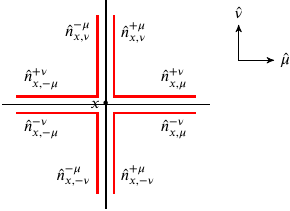}
  \caption{For \vconfiguration{}s we require two adjacent edge occupations of the same unoriented plaquette to be equal and call this requirement the hook condition. The quantity $\Delta_x$ in \eqref{eq:pc} is a product of hook conditions.}
  \label{fig:edge_occupations}
\end{figure}

The requirement for the edge variables to be equal around oriented plaquettes
translates into a condition for the edge variables corresponding to a given tensor $T_x$, see \cite[Eq.~(6.14)]{Samberger:2025hsr}. 
That condition implies a weaker condition for the edge occupations to obtain nonzero entries of $T_x$,
\begin{equation}
  \label{eq:pc}
  (T_x)_{\j_x} \propto \Delta_x
  \quad\text{with}\quad
  \Delta_x(j_x)\equiv
  \prod_{\mu,\nu=\pm1\atop|\mu|<|\nu|}^{\pm d}
  \delta_{\nn{\nu}_{x,\mu},\nn{\mu}_{x,\nu}} 
  \,,
\end{equation}
which is illustrated in \cref{fig:edge_occupations}. 
We will refer to the Kronecker deltas in \eqref{eq:pc} as hook conditions in the following.

The partition function also contains integrals over the auxiliary Grassmann variables $\aux_{x,\mu}$ which are contained in the local Grassmann tensor $K_x$, as shown in \cite[Sec.~5]{Samberger:2025hsr}. Each entry of the Grassmann tensor depends on Grassmann parities $\f_{x,\mu}(j_{x,\mu})\in\{0,1\}$, which indicate if the link has an even or odd Grassmann occupation.
We collect the Grassmann parities in the tuple $\f_x\equiv(\f_{x,-1},\f_{x,1},\dots, \f_{x,-d},\f_{x,d})$.

\subsection{Restriction of the edge variables}
\label{sec:trunc}

To contract the complete tensor network and compute the partition function and thermodynamical observables using the GHOTRG \cite{Bloch:2022vqz} method, the bond dimension of the initial tensor, which is the range of the compound link index $j_{x,\mu}$ in \eqref{eq:final_tensor_network},\footnote{For simplicity, we will often refer to the range of the link index $j_{x,\mu}$ as the bond dimension of the link $(x,\mu)$.}  has to be finite.
The compound link index contains variables of finite range, but also the edge variables, which run from zero to infinity as they originate from indices of a Taylor expansion, see \cite[Sec.~6.2]{Samberger:2025hsr}. To make the range of $j_{x,\mu}$ finite, we need to restrict the edge variables. 
To this end, we introduce a maximum order $\nmax$ in $\beta$ and remove all values of $j_{x,\mu}$
that do not satisfy the initial link criterion
\begin{equation}\label{eq:link_crit}
  \sum_{{\nu=\pm1}\atop{|\nu|\ne|\mu|}}^{\pm d}\nn{\nu}_{x,\mu}(j_{x,\mu})\leq \nmax\,,
\end{equation}
see \cite[Eq.~(6.17)]{Samberger:2025hsr} and \eqref{eq:def_plaq_excitation}.
With this restriction, all configurations of order $\beta^n$ with $0 \leq n \leq \nmax$ are included in the calculation of the partition function, while higher orders will be incomplete.  Without any TRG truncations, the partition function is then exact up to $\Order{\beta^{\nmax+1}}$ corrections.

Note that we can substantially speed up the tensor construction by setting to zero the entries of the initial tensor that do not satisfy the initial site criterion
\begin{equation}\label{eq:site_crit}
  n_x \le \frac14\nmax \,,
\end{equation}
see \cite[Eq.~(6.18)]{Samberger:2025hsr}.
These entries would correspond to configurations of order $\beta^n$ with $n>\nmax$. Although this step does not reduce the size of the initial tensor, it also increases the efficiency of the tensor-network method by removing singular values that would correspond to $n>\nmax$ and could be large.\footnote{After choosing $\nmax$, the bond dimension of the initial tensor is fixed. However, for an efficient numerical computation, especially for higher orders in $\beta$, the initial tensor may additionally be truncated to a particular bond dimension $D$, using an HOSVD-inspired truncation procedure analogous to that of the GHOTRG method.} 

A detailed derivation of more general versions of the link and site criteria, which are valid at every step of the blocking procedure, will be given in \cref{sec:reduction}.

\section{\boldmath Order separation in $\beta$: OS-GHOTRG method}
\label{Sec:Separation}

As discussed in the introduction, we would like to determine the expansion coefficients of $Z$ in $\beta$ up to a given order $\nmax$.
The order of a \vconfiguration $\bm{j}$ involves the whole lattice and cannot be determined by the local tensor alone. To keep track of the order of the contributions during the coarsening process and to remove higher-order contributions on the fly we developed a modified tensor renormalization group method, which we call order-separated GHOTRG (OS-GHOTRG) and describe in this section.

\subsection{Overview}

Let us briefly describe the main steps involved in computing the complete trace of the tensor network and define the corresponding terminology.
A blocking (or coarsening) step in direction $\mu$ starts from a ``fine'' grid and results in a ``coarse'' grid with half the number of lattice sites in direction $\mu$.
We divide a blocking step into three substeps: contraction, reduction, and truncation, as described in detail in the next three subsections. Here we only give a short qualitative outline.

For all even sites $x$ in direction $\mu$ on the fine grid, we take the tensors at $x$ and at the adjacent site $y=x+\hat\mu$ and contract them along the ``thin'' link $(x,\mu)$, which is shared by the two tensors. The sites $x$ and $y$ are fused into the site $X$ on the coarse grid, and every pair of thin links perpendicular to $\mu$ is merged into a ``fat'' link. We first contract the Grassmann tensors using GHOTRG, i.e., we integrate out the fine-grid Grassmann variables and introduce one new Grassmann variable per fat link \cite{Bloch:2022vqz,Milde2023}. 
This results in a new Grassmann tensor $K_X$ on the coarse grid.\footnote{To keep the notation simple we use the same symbol for a quantity on the fine and coarse grids.}
The reordering of Grassmann variables in this process produces sign factors, which are included in the numerical tensors.
We then contract the numerical tensors, resulting in a coarse-grid tensor $T_X$, where the fat links have squared bond dimensions. We will show that the entries of $T_X$ can be written as a power series in $\beta$.

After the contraction step, we perform a reduction step which removes ``invalid'' configurations (i.e., configurations whose contribution to $Z$ is zero) and contributions of $\mathcal{O}(\beta^{\nmax+1})$ from the partition function. This reduces the bond dimension of the fat link and sets some tensor entries to zero. 

Repeated application of these steps would still lead to an exponential increase of the size of the tensor.
To avoid this blowup, we perform a truncation step in which the bond dimension of the fat links is cut to a desired value. 
The truncation step differs from the standard GHOTRG truncation, as it is designed to preserve the structure of the tensor with entries that are power series in $\beta$.
This completes a blocking step.

\begin{figure}
\centering
\includegraphics[scale=1]{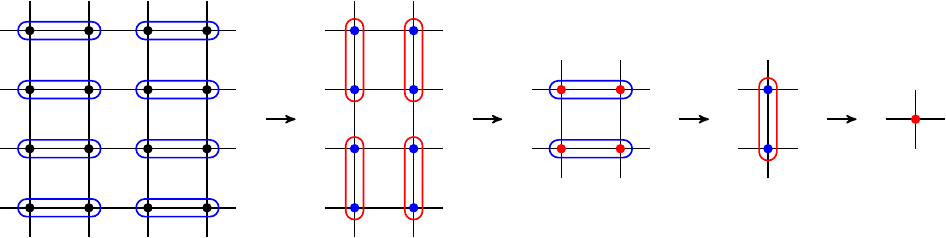}
\caption{Blocking procedure to reduce the tensor network on a two-dimensional $4\times 4$ lattice to a single tensor using alternating contractions in the horizontal and vertical directions. Each contraction step results in a new tensor network defined on a coarse grid with half the number of sites in the contraction direction and with fat links in the perpendicular directions, which are then truncated.}
\label{fig:RGblocking}
\end{figure}

The blocking steps are applied repeatedly in all directions\footnote{Once the blocking has been done at least once in each of the  directions $1,\ldots,d-1$, all staggered phases have been combined in $\Tnew$, which therefore no longer depends on its location on the lattice.\label{footnote:staggered}} until the final coarse grid has only one site, see \cref{fig:RGblocking} for a $4\times4$ lattice in two dimensions.
The partition function is then the trace of the tensor at this single site over all dimensions.
 
\subsection{Tensor contraction and decomposition}
\label{subsec:Occupation_during_contr}

\begin{figure}
  \centering
  \includegraphics{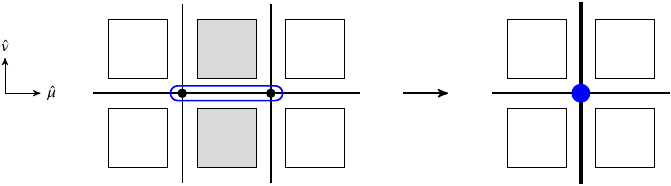}
  \caption{Illustration of a contraction step. On the fine grid (left), the tensors at the sites $x$ and $y=x+\hat\mu$ are contracted, resulting in a new tensor at the site $X$ on the coarse grid (right). Every plaquette that has the contracted link as an edge is collapsed onto the corresponding fat link perpendicular to $\hat\mu$.}
  \label{fig:coarsening}
\end{figure}

For the initial tensor, the $\beta$ dependence is given by \eqref{Tbetadep} and comes exclusively from the edge occupations.
We now perform a contraction step, in which we contract two adjacent tensors on the link between $x$ and $y=x+\hat\mu$. To understand what happens in this step we consider all plaquettes that have $x$ or $y$ as a corner and divide them into three groups: (1) plaquettes that have an edge along the contracted link and thus have both $x$ and $y$ as corners (indicated in gray in \cref{fig:coarsening}), (2) plaquettes that have either $x$ or $y$ as a corner and whose plane includes the contraction direction (indicated in white in \cref{fig:coarsening}), and (3) plaquettes in planes perpendicular to the contraction direction that have either $x$ or $y$ as corner (not shown in \cref{fig:coarsening}). 
Every plaquette in group (1) is collapsed onto the corresponding perpendicular fat link generated in the contraction, see \cref{fig:coarsening}. We shall see that the edge occupations of the collapsed plaquettes will be turned into a new link occupation $\nl_{X,\nu}\equiv\frac12(\nn{\mu}_{x,\nu} + \nn{-\mu}_{y,\nu})$ for every fat link $(X,\nu)$ perpendicular to $\hat\mu$, 
where positive or negative $\nu$ correspond to the upper and lower gray plaquette in \cref{fig:coarsening}, respectively. 
For plaquettes in group (2) the edge occupations on the coarse grid are the same as on the fine grid. 
Plaquettes in group (3), say in the plane $(\nu,\lambda)\perp\hat\mu$, are merged, and the edge occupations on the coarse grid are 
given by \eqref{eq:nhatXnulambda} below. 

In subsequent blocking steps the link occupations $\nl$ can increase further as more plaquettes in group (1) are collapsed onto fat links. 
Since links that are merged into fat links (perpendicular to the contraction direction) may already contain collapsed plaquettes, the formula for $\nl_{X,\nu}$ generalizes to \eqref{eq:nlXnu} below.

If the contracted link $(x,\mu)$ already contains a collapsed plaquette from a previous blocking step (i.e., $\nl_{x,\mu} \neq 0$), the contraction turns $\nl_{x,\mu}$ into a site occupation $\ns_X\equiv\nl_{x,\mu}$ of the entry of the coarse-grid tensor. If the sites $x$ and $y$ at the ends of the contracted link are already occupied through previous blocking steps, we obtain the more general formula \eqref{eq:nsX2} below.

In the following (including Secs.~\ref{sec:reduction} and \ref{sec:truncation}) we will show that, at every blocking step of the OS-GHOTRG method, the numerical tensor $T_x$ can be decomposed as
\begin{equation}\label{eq:Sum_split_of_tensor}
	(T_x)_{\j_x} = \beta^{n_x(\j_x)}\sum_{\ns_x=0}^\nmax \beta^{\ns_x}(T_x^{\ns_x})_{\j_x} \,,
\end{equation}
where the coefficient tensors $T_x^{\ns_x}$ are labeled by site occupations $\ns_x$, have the same bond dimensions as the full tensor $T_x$, are independent of $\beta$, and satisfy
\begin{equation}
  \label{eq:pcrg}
  (T_x^{\ns_x})_{\j_x}\propto\Delta_x(j_x)\quad\text{with $\Delta_x$ given in \eqref{eq:pc}}\,.
\end{equation}
The exponent $n_x$ of $\beta$ in \eqref{eq:Sum_split_of_tensor} is given by\footnote{For a \vconfiguration (satisfying the hook conditions) $n_x$ may not be an integer, but $\sum_xn_x$, which appears in the exponent of $\beta$ in $Z$, is.}
\begin{equation}  \label{eq:nx}
  n_x(\j_x)
  \equiv \frac12\sum_{\nu=\pm1}^{\pm d} \nl_{x,\nu}(j_{x,\nu})+\frac18\sum_{{\nu,\lambda=\pm1}\atop{|\nu|\ne|\lambda|}}^{\pm d}\nn{\lambda}_{x,\nu}(j_{x,\nu})\,,
\end{equation}
which generalizes the definition in \eqref{Tbetadep} to an arbitrary blocking step.
The link and edge occupations $\nl_{x,\mu}$ and $\nn{\nu}_{x,\mu}$ are determined by the tensor index $j_{x,\mu}$.

The initial tensor \eqref{Tbetadep} can be decomposed as in \eqref{eq:Sum_split_of_tensor} with
\begin{equation}
  \label{eq:Tx0}
	(T_x^0)_{j_x} \equiv \beta^{-n_x(j_x)} (T_x)_{j_x} 
	\quad \text{and} \quad (T_x^{\ns_x})_{j_x} \equiv 0 \quad \text{for } \ns_x\ge1 \,,
\end{equation}
where \eqref{eq:nx} holds with $\nl_{x,\mu}=0$.
From \eqref{eq:Tx0} and \eqref{eq:pc} it follows immediately that \eqref{eq:pcrg} holds for the initial tensor.
Furthermore, for the initial tensor $\nl_{x,\mu}$ and $\nn{\nu}_{x,\mu}$ are determined by $j_{x,\mu}$, the former trivially since it is zero, and the latter since the $\nn{\nu}_{x,\mu}$ are defined in terms of the edge variables, which in turn are part of $j_{x,\mu}$.

Now consider the contraction of two adjacent tensors on sites $x$ and $y = x+\hat\mu$ in the $\mu$ direction.
The contraction of the local Grassmann tensors $K_x$ and $K_y$ into a coarse-grid Grassmann tensor $K_X$ using the GHOTRG procedure is described in detail in \cite[Sec.~3]{Bloch:2022vqz} and \cite{Milde2023} for two and arbitrary dimensions, respectively.
The resulting sign factor $\sigma_{\f_x,\f_y}$ and the indices $F_X$ of $K_X$ are given in \cref{app:sign}.
The numerical tensor on the coarse grid, including the sign factor, is obtained as
\begin{equation}\label{eq:just_contraction}
  (\Tnew_X)_{\fatj_X} \equiv  \sum_{j_{x,\mu}}(T_x)_{\j_x}(T_y)_{\j_y}\sigma_{\f_x,\f_y}\,,
\end{equation}
where the contraction is performed by summing over the shared index $j_{x,\mu}=j_{y,-\mu}$. The coarse-grid tensor $\Tnew_X$ has fat indices
\begin{equation}\label{eq:fatlink}
  \fatj_{X,\nu}\equiv(j_{x,\nu},j_{y,\nu})\quad\text{for }|\nu|\ne\mu 
\end{equation}
defined on the fat links (perpendicular to the contraction direction), while the external indices in the contraction direction are $\fatj_{X,-\mu}\equiv j_{x,-\mu}$ and $\fatj_{X,\mu}\equiv j_{y,\mu}$. All these indices are combined in $\fatj_X$ as usual. 

\begin{figure}
  \centering
  \includegraphics{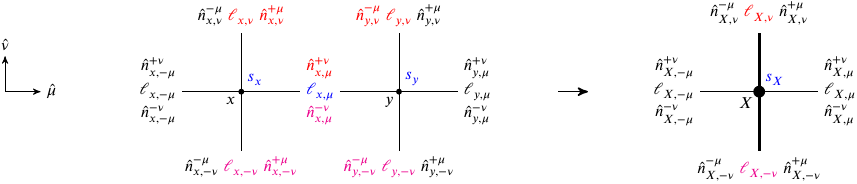}
  \caption{Visualization of the construction of the occupation numbers on a coarse site. The occupations on the coarse site (right) are computed from those on the fine sites (left) according to \eqref{eq:coarse_occupations}.}
  \label{fig:fat_gauge_occupations}
\end{figure}

Given the occupation numbers on the fine grid and assuming that $T_x$ and $T_y$ both satisfy \eqref{eq:Sum_split_of_tensor}, we show in \cref{app:induction} 
that \eqref{eq:Sum_split_of_tensor} also holds after the contraction for the coarse-grid tensor $\Tnew_X$ with
\begin{equation}
	(\Tnew_X^{\ns_X})_{\fatj_X} 
	=  
	\sum_{\ns_x=0}^{\ns_X}\sum_{\ns_y=0}^{\ns_X-\ns_x} \, 
	\sum_{j_{x,\mu}\atop\nl_{x,\mu}=\ns_X-\ns_x-\ns_y} 
	(T_x^{\ns_x})_{\j_x}
	(T_y^{\ns_y})_{\j_y}
    \sigma_{\f_x,\f_y}
	\label{eq:MnsX}
\end{equation}
and occupation numbers given by
\begin{subequations}
  \label{eq:coarse_occupations}
  \begin{align}
    \label{eq:nsX2}
    \ns_X&\equiv\ns_x + \ns_y + \nl_{x,\mu}\,,\\
    \label{eq:nlXnu}
    \nl_{X,\nu}&\equiv
    \begin{cases}
      \nl_{x,\nu} & \text{for }\nu=-\mu\,,\\
      \nl_{y,\nu} & \text{for }\nu=\mu\,,\\
      \nl_{x,\nu}+\nl_{y,\nu} + \frac12(\nn{\mu}_{x,\nu} + \nn{-\mu}_{y,\nu}) & \text{for }|\nu|\ne\mu\,,
    \end{cases}\\
    \label{eq:nhatXnulambda}
    \nn{\lambda}_{X,\nu}&\equiv
    \begin{cases}
      \nn{\lambda}_{x,\nu} & \text{for }\nu=-\mu\text{ or }\lambda=-\mu\,,\\
      \nn{\lambda}_{y,\nu} & \text{for }\nu=\mu\text{ or }\lambda=\mu\,,\\
      \nn{\lambda}_{x,\nu}+\nn{\lambda}_{y,\nu} & \text{for }|\lambda|,|\nu|\ne\mu\,,
    \end{cases}
  \end{align}
\end{subequations}
where in \eqref{eq:nhatXnulambda} we have $|\lambda|\ne|\nu|$.
This is visualized in \cref{fig:fat_gauge_occupations}. 
Note that $\nl_{X,\nu}$ and $\nn{\lambda}_{X,\nu}$ are determined uniquely by $\fatj_{X,\nu}$. 

\subsection{Reduction of the tensor}
\label{sec:reduction}

After a contraction the numerical tensor can be simplified in three different ways, which we now discuss in turn.
The first and second simplification reduce the size of the tensor by removing some of the fat-index values $\fatj_{X,\nu}=(j_{x,\nu},j_{y,\nu})$. 
This reduces the range of $\fatj_{X,\nu}$, i.e., the bond dimension of the fat link $(X,\nu)$. 
The second and third simplification remove information that contributes to $\mathcal O(\beta^{\nmax+1})$ in the partition function. All three simplifications preserve the structure of \eqref{eq:Sum_split_of_tensor}.

First, we consider the definition of $\nl_{X,\nu}$ in the third line of \eqref{eq:nlXnu}. 
The edge occupations $\nn{\mu}_{x,\nu}$ and $\nn{-\mu}_{y,\nu}$ are uniquely determined by $j_{x,\nu}$ and $j_{y,\nu}$, respectively. Although these edge occupations are independent variables for the tensors $T_x$ and $T_y$, the hook conditions in \eqref{eq:pc}, together with the link identity $(x,\mu)=(y,-\mu)$, imply that the entries of the coarse-grid tensor $\Tnew_X$ are zero if these two edge occupations are not equal.
Thus we can simply remove all values of the fat index $\fatj_{X,\nu}$ for which $\nn{\mu}_{x,\nu} \neq\nn{-\mu}_{y,\nu}$ without losing information.

Second, we remove all fat-index values $\fatj_{X,\nu}$ for which the link criterion 
\begin{equation}
  \label{eq:cutoff}
  \nl_{X,\nu}+\sum_{{\lambda=\pm1}\atop{|\lambda|\ne|\nu|}}^{\pm d}\nn{\lambda}_{X,\nu}\le\nmax
\end{equation}
is not satisfied. 
Third, we set all entries of the coefficient tensors to zero for which the site criterion\footnote{Note that link and site criterion have to be slightly altered when the coarse grid only contains a single site in a particular direction, as the backward and forward links are identical due to the toroidal shape of the grid. This case can be taken into account by including a factor of $t_\lambda$ (or $t_\nu$) for every summation index $\lambda$ (or $\nu$) in \eqref{eq:cutoff} and \eqref{eq:barnX}, where $t_\lambda=1/2$ if the lattice extent in the $\lambda$ direction is one and $t_\lambda=1$ otherwise (and similarly for $t_\nu$).}
\begin{equation}
  \label{eq:barnX}
  \ns_X+\sum_{\nu=\pm1}^{\pm d} \nl_{X,\nu}+\sum\limits_{\nu,\lambda=\pm1\atop|\nu|<|\lambda|}^{\pm d} \nn{\lambda}_{X,\nu}\le\nmax
\end{equation}
is not satisfied. The link and site criteria, which remove $\mathcal O(\beta^{\nmax+1})$ contributions to the partition function, are derived in \cref{app:link-site}.

In the actual implementation of the OS-GHOTRG method the reductions described above are applied on the fly during the contraction performed in \cref{subsec:Occupation_during_contr}.

\subsection{Truncation of the tensor}\label{sec:truncation}

When computing the partition function \eqref{eq:final_tensor_network} numerically using the GHOTRG method, we contract pairs of adjacent tensors at each blocking step, thus halving the number of tensors on the grid. This is repeated until the tensor network has been reduced to a single tensor, which is then traced over all directions to yield the partition function. If we were to contract and reduce tensors as detailed in \cref{subsec:Occupation_during_contr,sec:reduction}, the size of the tensor would increase exponentially during the blocking procedure. A crucial ingredient of the GHOTRG method is the truncation of the bond dimension of the fat links (perpendicular to the contraction direction) after each contraction step to avoid such an exponential blowup. The truncation is based on the HOSVD concept to reduce the multi-rank of a tensor  \cite{DeLathauwer2000}, in which an approximation $\hat T$ to a tensor $T$ is constructed as
\begin{equation}\label{eq:hosvd}
  \hat T_{i_1i_2\cdots}=U^{(1)}_{i_1j_1}U^{(2)}_{i_2j_2}\cdots \Ttrunc_{j_1j_2\cdots}\quad\text{with}\quad
  \Ttrunc_{j_1j_2\cdots}=U^{(1)}_{i_1j_1}U^{(2)}_{i_2j_2}\cdots T_{i_1i_2\cdots}\,,
\end{equation}
where typically $\range(j_k)\ll\range(i_k)$. Here, sums over repeated indices are implied, $\Ttrunc$ is called the core tensor, and the $U^{(k)}$ are tall and skinny semi-orthogonal matrices. 
The core tensor $\Ttrunc$ has smaller multi-rank and smaller bond dimensions than $T$. The tensor approximation $\hat T$ has the same multi-rank as $\Ttrunc$ but the same bond dimensions as $T$.

We now apply this concept to the numerical part $\Tnew_X$ of the coarse-grid tensor, while the Grassmann part is briefly discussed at the end of this subsection.
As in \eqref{eq:hosvd} we construct a core tensor
\begin{equation}\label{eq:truncT}
	(\Ttrunc_X)_{\jtrunc_X}\equiv
    \Biggl(\prod_{\nu=\pm1\atop|\nu|\ne\mu}^{\pm d}\sum_{\fatj_{X,\nu}}\Biggr)
    \Biggl(\prod_{\nu=\pm1\atop|\nu|\ne\mu}^{\pm d}(U_{X,\nu})_{\fatj_{X,\nu},\jtrunc_{X,\nu}}\Biggr)(\Tnew_X)_{\fatj_X}\,,
\end{equation}
where $\jtrunc_X\equiv(\jtrunc_{X,-1},\jtrunc_{X,1},\dots, \jtrunc_{X,-d},\jtrunc_{X,d})$ with $\jtrunc_{X,\pm\mu}=j_{X,\pm\mu}$ for the contraction direction $\mu$,
and $U_{X,\nu}$ is a semi-orthogonal matrix whose columns are typically singular vectors of a matrix constructed using tensor unfoldings (a.k.a.\ matricizations) of the coarse-grid tensor with respect to its fat indices, see for example \cite{Bloch:2022qyw}. 
The first dimension of $U_{X,\nu}$ is the original bond dimension of the fat link, and the second dimension is the desired bond dimension after truncation. We will refer to the $U_{X,\nu}$ as truncation matrices. 

In HOSVD, the tensor $T$ would be approximated by the tensor $\hat T$ with reduced multi-rank.
However, it is straightforward to show, using the semi-orthogonality properties of the truncation matrices, that the partition function is unchanged if we simply use the core tensor $\Ttrunc$ instead of $\hat T$, provided that the condition
\begin{equation}\label{eq:U}
  U_{X,\nu} = U_{X+\hat\nu,-\nu}
\end{equation}
is imposed on the truncation matrices.
Hence the core tensor $\Ttrunc_X$, which has (much) smaller bond dimensions than $\hat T_X$, is the truncated tensor in GHOTRG.

Inserting the decomposition of the coarse-grid tensor $T_X$ (obtained from \eqref{eq:Sum_split_of_tensor} with $x\to X$ and reduced as explained in \cref{sec:reduction}) in \eqref{eq:truncT}  yields
\begin{equation}
	(\Ttrunc_X)_{\jtrunc_X}
	= \sum_{\ns_X=0}^{\nmax} \beta^{\ns_X}\Biggl(\prod_{\nu=\pm1\atop|\nu|\ne\mu}^{\pm d}\sum_{\fatj_{X,\nu}}\Biggr)
    \Biggl(\prod_{\nu=\pm1\atop|\nu|\ne\mu}^{\pm d}(U_{X,\nu})_{\fatj_{X,\nu},\jtrunc_{X,\nu}}\Biggr) \beta^{n_X} (\Tnew_X^{\ns_X})_{\fatj_X}
	\,.
	\label{eq:truncTdecomp}
\end{equation}
The exponent $n_X$ depends on the $\fatj_{X,\nu}$. 
Due to the sum over $\fatj_{X,\nu}$ in \eqref{eq:truncTdecomp}, a term in the sum over $\ns_X$ contains contributions of different orders in $\beta$, i.e., these contributions mix. 
As we eventually want to obtain a decomposition of the partition function in powers of $\beta$ we will modify the GHOTRG method
so that the structure of \eqref{eq:Sum_split_of_tensor} is  preserved after truncation, i.e.,
\begin{equation}
  \label{eq:Tnewtrunc}
	(\Ttrunc_X)_{\jtrunc_X}
	= \beta^{n_X} \sum_{\ns_X=0}^{\nmax} \beta^{\ns_X} (\Ttrunc_X^{\ns_X})_{\jtrunc_X}
\end{equation}
with coefficient tensors
\begin{equation}\label{eq:TnsXtrunc}
  (\Ttrunc_X^{\ns_X})_{\jtrunc_X} = \Biggl(\prod_{\nu=\pm1\atop|\nu|\ne\mu}^{\pm d}\sum_{\fatj_{X,\nu}}\Biggr)
  \Biggl(\prod_{\nu=\pm1\atop|\nu|\ne\mu}^{\pm d}(U_{X,\nu})_{\fatj_{X,\nu},\jtrunc_{X,\nu}}\Biggr) (\Tnew_X^{\ns_X})_{\fatj_X} \,.
\end{equation}
This can be achieved by ensuring that the truncated indices $\jtrunc_{X,\nu}$ again uniquely determine the occupation numbers $\nl_{X,\nu}$ and $\nn{\mu}_{X,\nu}$.
In the following we explain this central part of the OS-GHOTRG method.

We start with the unfolding of the coarse-grid tensor $\Tnew_X$ with respect to the fat index  $\fatj_{X,\nu}$,\footnote{Note that the OS-GHOTRG method explicitly produces the coefficient tensors $\Tnew_X^{\ns_X}$. Therefore, to numerically compute the tensor entry $(\Tnew_X)_{\fatj_X}$ using \eqref{eq:Sum_split_of_tensor} an explicit choice of $\beta$ is required. This choice could be viewed as an optimization opportunity, but for the numerical results of \cref{sec:results} we always used $\beta=1$.}
\begin{equation}\label{eq:unfolding}
	(M_{X,\nu})_{\fatj_{X,\nu},\fatj_X\backslash \fatj_{X,\nu}}
	\equiv (\Tnew_X)_{\fatj_X}\,, 
\end{equation}
where $\fatj_X\backslash \fatj_{X,\nu}$ stands for the index tuple $\fatj_X$ without the index $\fatj_{X,\nu}$. 
The index $\fatj_{X,\nu}$ labels the rows of $M_{X,\nu}$, while the combination of all remaining indices labels its columns. 

We first show that the edge occupations are not mixed in \eqref{eq:truncTdecomp}. 
The index $\fatj_{X,\nu}$ uniquely determines all edge occupations $\nn{\lambda}_{X,\nu}$ (with $|\lambda|\neq|\nu|$). For nonzero entries the edge occupations satisfy the hook condition $\nn{\lambda}_{X,\nu}=\nn{\nu}_{X,\lambda}$, and hence the nonzero entries of a row of $M_{X,\nu}$ will only occur for columns for which the hook conditions are satisfied, i.e., all $\nn{\nu}_{X,\lambda}$ in the combined column index must match the $\nn{\lambda}_{X,\nu}$ of the row index $\fatj_{X,\nu}$. This automatically yields a block-diagonal structure in the unfolding.\footnote{A permutation of rows and columns may be necessary to make the block-diagonal structure explicit.} An SVD of the unfolding respects this blocking, i.e., the singular vectors\footnote{We always consider the left singular vectors, which form a square matrix whose dimension equals the row dimension of the unfolding.} have the same block structure as the unfolding itself, and thus each singular vector, labeled by $\jtrunc_{X,\nu}$, can be uniquely allocated values of $\nn{\lambda}_{X,\nu}$ for all $\lambda$ with $|\lambda|\neq|\nu|$.

We now turn to the link occupations, which are affected by mixing. The fat index $\fatj_{X,\nu}$ also determines the link occupation $\nl_{X,\nu}$. When computing the SVD and applying the truncation matrix, the different $\nl$ values will mix such that the truncated index $\jtrunc_{X,\nu}$ no longer uniquely determines the link occupation and the decomposition \eqref{eq:Tnewtrunc} does not hold. We now introduce a procedure to construct truncation matrices $U_{X,\nu}$ for which the truncated index $\jtrunc_{X,\nu}$ still uniquely determines the link occupation $\nl_{X,\nu}$.

To avoid the mixing of different link occupations $\nl_{X,\nu}$ under truncation, we perform separate SVDs of sub-unfoldings with fixed link occupation. To this end we reorder the rows, labeled by the index $\fatj_{X,\nu}$, in increasing order of link occupation $\nl_{X,\nu}(\fatj_{X,\nu})$ and obtain
\begin{equation}\label{eq:blockM}
	M_{X,\nu}
	=
	\begin{pmatrix}
		M_{X,\nu}^0\\[1mm] 
		M_{X,\nu}^1\\[1mm]
		\vdots \\[1mm]
		M_{X,\nu}^\nmax
	\end{pmatrix}
	\,,
\end{equation}
where each block $M_{X,\nu}^m$ corresponds to a fixed value $\nl_{X,\nu}=m$. Provided that \eqref{eq:U} has been satisfied in all preceding blocking steps, $M_{X,\nu}$ and $M_{X+\hat\nu,-\nu}$ have the same vertical block structure as in \eqref{eq:blockM}.

To obtain singular vectors that do not mix link occupations and to construct the truncation matrix $U_{X,\nu}$,
we apply one of two procedures, a symmetric one based on the SuperQ method \cite{Bloch:2022qyw} or an asymmetric one based on the original idea by Xie et al. \cite{Xie_2012}, where symmetric and asymmetric refers to the two tensors at the ends of the truncated link.
In SuperQ, we perform SVDs of the $\nmax+1$ extended matrices
\begin{equation}
  \label{eq:M_extended}
	\begin{pmatrix}
		M_{X,\nu}^m & M_{X+\hat\nu,-\nu}^m
	\end{pmatrix} \quad\text{with}\quad 0\le m\le\nmax\,.
\end{equation}
The resulting singular-vector matrices $U_{X,\nu}^m$ are assembled in the block-diagonal square matrix
\begin{equation}\label{eq:Ubar}
	\bar U_{X,\nu} =
	\begin{pmatrix}
		U_{X,\nu}^0 & 0 & \dots & 0 \\
		0 & U_{X,\nu}^1 & \dots & 0 \\
		\vdots & \vdots & \ddots & \vdots \\
		0 & 0 & \dots & U_{X,\nu}^\nmax
	\end{pmatrix}\,.
\end{equation}
The semi-orthogonal truncation matrix $U_{X,\nu}$ is then obtained by retaining the columns of $\bar U_{X,\nu}$ corresponding to the $D$ largest singular values of the set of matrices in \eqref{eq:M_extended}, where $D$ is the desired bond dimension of the truncated coarse-grid tensor.
In the Xie et al. procedure, we perform SVDs of the matrices $M_{X,\nu}^m$ and $M_{X+\hat\nu,-\nu}^m$ separately, again for $0\le m\le\nmax$, resulting in two matrices $\bar U_+$ and $\bar U_-$ constructed as in \eqref{eq:Ubar}, from which we obtain $U_+$ and $U_-$ as before. For $U_{X,\nu}$ we then pick either $U_+$ or $U_{-}$ depending on which gives the smaller truncation error for $M_{X,\nu}$ or $M_{X+\hat\nu,-\nu}$, respectively.\footnote{The truncation error is $1-\sum_{i=1}^D\sigma_i^2/\sum_{i=1}^{N}\sigma_i^2$, where the $\sigma_i$ are the singular values and $N$ is the dimension of the fat index. Usually $N=D^2$.}
SuperQ is preferred since it results in a smaller combined truncation error for $M_{X,\nu}$ and $M_{X+\hat\nu,-\nu}$ \cite{Bloch:2022qyw}. If it becomes numerically unstable, which can happen close to the phase transition for large volume, we use the Xie et al. procedure instead.

The truncation matrix $U_{X,\nu}$ is block-diagonal, with blocks corresponding to different $\nl_{X,\nu}$.
To apply $U_{X,\nu}$ in \eqref{eq:truncTdecomp} its rows are sorted back to the original order of the fat index $\fatj_{X,\nu}$.
The truncated index $\jtrunc_{X,\nu}$ corresponds to the columns of $U_{X,\nu}$ and uniquely determines $\nl_{X,\nu}$.
Together with the block structure discussed above for $\nn{\lambda}_{X,\nu}$ this ensures that the decomposition \eqref{eq:Tnewtrunc} holds, with the coefficient tensors given by \eqref{eq:TnsXtrunc}.

We now turn to the Grassmann part of the coarse-grid tensor.
From \cite{Bloch:2022vqz,Milde2023} we know that $M_{X,\nu}$ is block-diagonal in the Grassmann parity.
This Grassmann-parity-based block structure also holds for the individual $M_{X,\nu}^m$ blocks.
This implies that the matrices $U_{X,\nu}^m$ are themselves block-diagonal, where each of the two diagonal blocks corresponds to a different Grassmann parity, see \cite[Sec.~3.2.3]{Bloch:2022vqz}. If we apply the truncation matrices to the product of $T_X$ and $K_X$, the ``truncated'' Grassmann tensor is simply $K_X$ with $\f_{X,\nu}(j_{X,\nu})$ replaced by $\bar\f_{X,\nu}(\bar j_{X,\nu})$, which is the Grassmann parity of column $\bar j_{X,\nu}$ of $U_{X,\nu}$. For details see \cite[Sec.\ 3.2.4]{Bloch:2022vqz}.

At the end of every blocking step, we rename $X\to x$ and drop the bar on all quantities.
The partition function then has the form of \eqref{eq:final_tensor_network} at every step of the blocking procedure, with half the number of lattice sites and new numerical tensors.
As mentioned in \cref{footnote:staggered}, $T_X$ will be independent of $X$ after a sufficient number of blocking steps (at least $d-1$ steps). This implies that the complexity of the tensor-network method scales like $\log V$.

\subsection{Tensor trace}

When the lattice is reduced to a single site\footnote{We assume that the linear extent of the lattice is a power of $2$ in every direction.} in a certain direction, say $\mu$, during the iterative blocking procedure, the tensor will be traced in that direction and the dimension of the tensor network is reduced from $d$ to $d-1$, where $d$ is the current dimension of the network.
In this process we again have to separate contributions from different orders in $\beta$.
This is the subject of this subsection.

As part of the GHOTRG procedure \cite{Bloch:2022vqz}, we integrate out the remaining Grassmann variable in direction $\mu$ and apply the corresponding boundary condition, resulting in a sign factor
\begin{equation}\label{eq:sigma_mu}
  \sigma_\mu=(-1)^{\f_{x,\mu}[1+\delta_{\mu,\text{time}} +\sum_{\nu=\mu+1}^d(\f_{x,\nu}+\f_{x,-\nu})]}\,.
\end{equation}
We then compute the trace to obtain
\begin{equation}\label{eq:trace}
  (T_X)_{j_X}\equiv\sum_{j_{x,\mu}} (T_x)_{j_x}\sigma_{\mu}
\end{equation}
with summation index $j_{x,\mu}=j_{x,-\mu}$. Here, $X$ is the site on the $(d-1)$-dimensional lattice obtained from $x$ by dropping the coordinate $x_\mu$, and $j_X\equiv j_x\backslash \,(j_{x,-\mu},j_{x,\mu})$.

Given the occupation numbers on the $d$-dimensional lattice and assuming that $T_x$ satisfies \eqref{eq:Sum_split_of_tensor}, we show in \cref{app:trace} that \eqref{eq:Sum_split_of_tensor} also holds after the tracing for $T_X$ on the $(d-1)$-dimensional lattice, with 
\begin{equation}\label{eq:TXsXd-1}
  (T_X^{\ns_X})_{j_X}=\sum_{\ns_x=0}^{\ns_X}\sum_{j_{x,\mu}\atop{\nl_{x,\mu}=\ns_X-\ns_x}}(T_x^{\ns_x})_{j_x}\sigma_{\mu} \,,
\end{equation}
occupation numbers for $|\nu|,|\lambda|\ne\mu$ and $|\nu|\ne|\lambda|$ given by
\begin{subequations}\label{eq:occupations_after_tracing}
  \begin{align}
    s_X&=s_x+\nl_{x,\mu}\,,\\
    \nl_{X,\nu}&=\nl_{x,\nu}+\frac12(\nn{\mu}_{x,\nu}+\nn{-\mu}_{x,\nu})\,,\\
    \nn{\lambda}_{X,\nu}&=\nn{\lambda}_{x,\nu}\,,
  \end{align}
\end{subequations}
and
\begin{equation}\label{eq:nXd-1}
  n_X(j_X)= \frac12 \sum_{\nu=\pm1\atop{|\nu|\ne\mu}}^{\pm d} \nl_{X,\nu} + \frac18 \sum_{\nu,\lambda=\pm1\atop{{|\nu|,|\lambda|\ne\mu}\atop{|\nu|\ne|\lambda|}}}^{\pm d} \nn{\lambda}_{X,\nu}\,.
\end{equation}
Note that $\nl_{X,\nu}$ and $\nn{\lambda}_{X,\nu}$ are uniquely determined by $j_{X,\nu}$. 

We now rename $X\to x$ and relabel the $d-1$ remaining directions from $1$ to $d-1$ to be able to start the next blocking step using the same notation as before. Note that once we have arrived at a one-dimensional lattice, the blocking no longer involves a truncation since there is no perpendicular direction.

Finally, when the one-dimensional lattice has been reduced to a single site, we use \eqref{eq:trace} to compute the partition function  \eqref{eq:final_tensor_network} up to order $\nmax$,
\begin{equation}
  Z=\sum_{j_{x,1}}(T_x)_{j_{x,-1},j_{x,1}}\sigma_1
\end{equation}
with $j_{x,-1}=j_{x,1}$ and $\sigma_1$ from \eqref{eq:sigma_mu}. Using the decomposition \eqref{eq:Sum_split_of_tensor} with \eqref{eq:TXsXd-1} and \eqref{eq:occupations_after_tracing}, now for $d=1$, we obtain the partition function as a polynomial in $\beta$, 
\begin{equation}\label{eq:Zsce}
  Z=\sum_{n=0}^{\nmax} Z_{n} \beta^{n}
  \quad\text{with}\quad
  Z_{n} = \sum_{\ns_x=0}^{n}
  \sum_{j_{x,1}\atop{\nl_{x,1}=n-\ns_x}}
  (T_x^{\ns_x})_{j_{x,1},j_{x,1}}\sigma_1\,.
\end{equation}

\section{Translational invariance}
\label{sec:trans_inv}

For $\nmax\ge1$ we can increase the numerical efficiency by employing translational invariance on the initial lattice. 
Configurations that only differ by a translation on the lattice make the same contribution to the partition function.
Therefore, we would like to include only a subset of these configurations explicitly and apply a combinatorial factor to also take into account the others. 
For simplicity, we only discuss this idea for the two-dimensional case. 
An extension to more than two dimensions is the subject of future work.

\subsection{Contributions to the partition function}
\label{sec:contributions}

Let $(q_1,\ldots,q_r)$ be a tuple of $r$ positive integers.
We define $Z_{(q_1,\ldots,q_r)}$ to be the sum of all contributions to the partition function of \vconfiguration{}s $\bm j$ with nonzero plaquette occupations\footnote{For a \vconfiguration the hook conditions are satisfied and the plaquette occupation is the same as the four identical edge occupations around this plaquette, see \cref{fig:plaq_to_link}. When we speak of plaquette occupations we always mean those of the initial lattice.} as in the tuple.
Furthermore, we define $Z_{(q_1|q_2,\dots,q_r)}$ just like $Z_{(q_1,\ldots,q_r)}$, but with the additional constraint that the plaquette at the origin has occupation $q_1$.

In $Z_{(q_1,\ldots,q_r)}$ the order of the $q_i$ is irrelevant since  we sum over all possible locations of the plaquette occupations on the lattice. Hence, to obtain all contributions to $Z_n$ in \eqref{eq:Zsce} it is sufficient to sum $Z_{(n)}$ over all partitions $(n)$ of $n$, defined by\footnote{In the interest of readability we have overloaded the notation, denoting by $(n)$ both a generic partition and the specific partition with $n_1=n$.} $(n)=(n_1,\ldots,n_1,n_2,\ldots,n_2,\ldots,n_m,\ldots,n_m)$, where $n_1>n_2>\ldots>n_m>0$ and $k_i$ is the multiplicity of $n_i$, i.e., $n=\sum_{i=1}^mk_in_i$.
Translational invariance allows us to obtain $Z_{(n)}$ as\footnote{Translational invariance implies that the contributions of two \vconfiguration{}s $\bm{j}$ and $\bm{j}'$ with $j'_{x} = j_{x-y}$ are equal in \eqref{eq:final_tensor_network}, where $y$ is an arbitrary shift on the toroidal lattice.
An explicit analysis \cite{Samberger2022}, taking into account the staggered phases and boundary conditions, confirms that this is indeed the case. 
Now consider a \vconfiguration $\bm j$ with nonzero plaquette occupations at lattice sites $x_1,\ldots,x_r$ given as in $(n)$, where $r=\sum_{i=1}^mk_i$.
There are $V$ shifted replicas $\bm j'$ of this configuration (including shift $y=0$) which make identical contributions to $Z_{(n)}$.
Out of these $V$ replicas, only those $k_i$ replicas are kept in $Z_{(n_i|\,n\backslash n_i)}$ for which the shift $y$ is such that the occupation of the plaquette at the origin is $n_i$. If $(x_1,\dots,x_r)$ has special symmetries such that some shifted configurations are identical to one another on the torus (e.g., if for a lattice of size $L\times L$ and partition $(1,1)$ with $x_1=(0,0)$ and $x_2=(L/2,L/2)$ we shift by $y=(L/2,L/2)$), the number of replicas is reduced to $V/s$ and $k_i/s$, respectively, with symmetry factor $s$. 
Therefore all configurations considered in $Z_{(n_i|\,n\backslash n_i)}$ contribute to $Z_{(n)}$ with combinatorial factor $V/k_i$ independent of $x_1,\ldots,x_r$.\label{footnote:translational_invariance}}
\begin{equation}\label{eq:Zn}
  Z_{(n)}=\frac V{k_i}Z_{(n_i|\,n\backslash n_i)}\quad\text{for any }i\in\{1,\ldots,m\}\,,
\end{equation}
where $(n\backslash n_i)$ means that the first occurrence of $n_i$ in $(n)$ has been removed.
The desired coefficients $Z_n$ are thus given by
\begin{subequations}\label{eq:Zi_Z(i)}
  \begin{align}
    Z_1 &= Z_{(1)} = V Z_{(1|)} \,, \\  
    Z_2 &= Z_{(2)} + Z_{(1,1)} = V Z_{(2|)} + \frac V2 Z_{(1|1)} \,, \\ 
    Z_3 &= Z_{(3)} + Z_{(2,1)} + Z_{(1,1,1)} = V Z_{(3|)} + V Z_{(2|1)} + \frac V3 Z_{(1|1,1)} \,,\\
    Z_4 &= Z_{(4)} + Z_{(3,1)} + Z_{(2,2)} + Z_{(2,1,1)} + Z_{(1,1,1,1)}
    = V Z_{(4|)} + V Z_{(3|1)} + \frac V2 Z_{(2|2)} + V Z_{(2|1,1)} + \frac V4 Z_{(1|1,1,1)}
  \end{align}
\end{subequations}
and so on for larger $n$. 
We now construct two tensor networks from which we will determine the $Z_n$.\footnote{It is conceivable that translational invariance could be exploited in a slightly different way using more than two tensor networks, where in each network the occupation of the origin plaquette is fixed at a particular value. The pros and cons of such an approach have not yet been explored.}

\subsection[$\X$ network]{\boldmath $\X$ network}
\label{sec:tildeZ}

\begin{figure}
  \centering
  \includegraphics{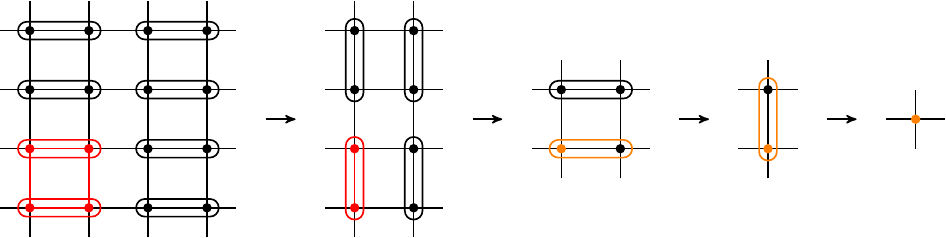}
  \caption{Illustration of the blocking procedure using translational invariance on the initial lattice. The plaquette at the origin and the $S$ tensors are indicated in red, the $\tilde S$ tensors are indicated in orange, and restricted plaquettes and $R$ tensors are indicated in black, respectively.}
\label{fig:Iterative_contraction_with_plaq_at_origin}
\end{figure}

For the first tensor network, which we call the $\X$ network, we require the occupation of the plaquette at the origin to be nonzero. To this end we introduce four initial ``special'' tensors $S_1,\dots,S_4$ at the corners of the plaquette at the origin.
We then restrict the occupations of all other plaquettes, which we call restricted plaquettes, to $n\leq \nR$. Here, $\nR$ is a parameter in the interval $[\lfloor\nmax/2\rfloor,\nmax-1]$,\footnote{$\lfloor\cdots\rfloor$ denotes the floor function.} where the lower bound ensures that all configurations can be included and the upper bound follows from the fact that the occupation of the plaquette at the origin is at least one.
The tensors $S_i$ are surrounded by the plaquette at the origin with plaquette occupation $n\in\{1,\dots,\nmax\}$ (red plaquette in \cref{fig:Iterative_contraction_with_plaq_at_origin}) and three restricted plaquettes with occupation $n\in\{0,\dots,\nR\}$ (black plaquettes). The tensors on all other lattice sites, which we call ``restricted'' tensors $R$, are surrounded by four restricted plaquettes.\footnote{Both the $S$ and $R$ tensors have reduced dimensions as some or all of the surrounding plaquettes are restricted. Also, both tensor types still satisfy the initial link and site criteria \eqref{eq:link_crit} and \eqref{eq:site_crit}.} 
 
The blocking procedure now proceeds as shown in \cref{fig:Iterative_contraction_with_plaq_at_origin}.
After one blocking step in each of the two directions, the volume $V$ has been reduced by a factor of 4, and the occupation of the plaquette at the origin has been turned into the site occupation of the resulting $S$ tensor. We decompose this tensor as in \eqref{eq:Sum_split_of_tensor} but introduce additional scaling coefficients $c_s$, resulting in the tensor
\begin{equation}\label{eq:Stilde}
(\tilde S)_{\j} \equiv \beta^{n(\j)}\sum_{\ns=1}^\nmax \beta^{\ns}(\tilde S^{\ns})_{\j} 
\quad \text{with} \quad \tilde S^{\ns} \equiv c_{\ns} S^{\ns} \,,
\end{equation}
where the $c_s$ are determined below. Note that $s$ corresponds to the occupation of the plaquette at the origin of the initial lattice 
and that the sum starts at $s=1$ since here we only consider configurations for which this occupation is at least one.

Denoting a blocking step in direction $\mu$ by $\star_\mu$, the steps can be summarized as
\begin{equation}\label{eq:Xblocking}
  \begin{aligned}
    &\text{Step $1$:} & R^{(0)} \star_1 R^{(0)} &\to R^{(1)}, &  S^{(0)}_1 \star_1 S^{(0)}_2 &\to S^{(1)}_1, \qquad S^{(0)}_{3} \star_1 S^{(0)}_{4} \to S^{(1)}_2 ,\\
    &\text{Step $2$:} & R^{(1)} \star_2 R^{(1)} &\to R^{(2)}, & S^{(1)}_1 \star_2 S^{(1)}_2 &\to S^{(2)} \to \tilde S^{(2)} \quad\text{using \eqref{eq:Stilde}}\,,\\
    &\text{Step $i+1$:} & R^{(i)} \star_\mu R^{(i)} &\to R^{(i+1)}, & \tilde S^{(i)} \star_\mu R^{(i)} &\to \tilde S^{(i+1)} \quad \text{for } i\ge2 \,,
  \end{aligned}
\end{equation}
where the last line is repeated in all directions until we arrive at a single point.\footnote{Because of the contractions between $\tilde S$ and $R$ we have to revisit the truncation procedure. 
If we were to apply the truncation procedure of \cref{sec:truncation}, the dependence of the tensor on the location $x$ would no longer disappear, which would destroy the $\log V$ scaling of the tensor-network method. 
We therefore construct the truncation matrices only from the $R$ tensors, except for the truncation of $S_1^{(1)}$ and $S_2^{(1)}$.}

With this modified blocking procedure we now compute the full contraction of the tensor network involving one tensor $\tilde S$ and $V/4-1$ restricted tensors $R$, written symbolically as
\begin{equation}\label{eq:Ztilde}
  \X = \tr \left[\tilde S^{(2)} \left(R^{(2)}\right)^{V/4-1}\right],
\end{equation}
using the OS-GHOTRG method of Sec.~\ref{Sec:Separation}.
In the reduction step of the OS-GHOTRG method, see \cref{sec:reduction}, we can replace $\nmax$ by $\nmax-1$ when computing $R \star R$ since the occupation of the plaquette at the origin is nonzero. This reduces the numerical effort significantly. 
From \eqref{eq:Zsce} we obtain
\begin{equation}
  \X = \sum_{n=1}^\nmax \X_n \beta^n
\end{equation}
with
\begin{subequations}
  \begin{align}
    \X_1 &= c_1 Z_{(1|)} \,, \\
    \X_2 &= c_2 Z_{(2|)} + c_1 Z_{(1|1)} \,, \\
    \X_3 &= 
      c_3 Z_{(3|)} + c_2 Z_{(2|1)} + c_1 \bigl( \Theta(\nR-2) Z_{(1|2)} + Z_{(1|1,1)} \bigr) \notag\\
    &= 
      c_3 Z_{(3|)} + \bigl(c_2+c_1\Theta(\nR-2)\bigr) Z_{(2|1)} + c_1 Z_{(1|1,1)} \,, \label{eq:X3}
	\\
    \X_4 &= 
      c_4 Z_{(4|)} + c_3 Z_{(3|1)} + c_2 \bigl( Z_{(2|2)} + Z_{(2|1,1)} \bigr) + c_1 \bigl( \Theta(\nR-3) Z_{(1|3)} + Z_{(1|2,1)} + Z_{(1|1,1,1)} \bigr) \notag\\
      &= c_4 Z_{(4|)} + (c_3+c_1\Theta(\nR-3)) Z_{(3|1)} + c_2 Z_{(2|2)} + (c_2+2c_1) Z_{(2|1,1)} + c_1 Z_{(1|1,1,1)} \label{eq:X4}
  \end{align}
\end{subequations}
and so on for larger $n$. Here, $\Theta(x)$ is the Heaviside function with $\Theta(0)\equiv1$. To go from the first to the second line in \eqref{eq:X3} and \eqref{eq:X4} we shifted the configurations such that the highest plaquette occupation is at the origin and used
\begin{equation}\label{eq:Zperm}
Z_{(n_i | n \backslash n_i)} = \frac{k_i}{k_1} Z_{(n_1 | n \backslash n_1)}\,,
\end{equation}
which follows from \eqref{eq:Zn}.

\subsection[$\Y$ network]{\boldmath $\Y$ network}

We also compute the full contraction of another tensor network consisting only of tensors $R$, written symbolically as 
\begin{equation}
  \Y = \tr\left[\left(R^{(0)}\right)^{V}\right] = 
  \tr\left[\left(R^{(2)}\right)^{V/4}\right].
\end{equation}
Note that $\X$ and $\Y$ can be computed together efficiently, since $R\star R$ is needed for both quantities. Applying the OS-GHOTRG method with restriction to order $\nmax-1$ in each blocking step, we obtain
\begin{equation}
  \Y = \sum_{n=0}^{\nmax-1} \Y_n \beta^n \,,
\end{equation}
where the $\Y_n$ contain the contributions to $Z_n$ of all configurations for which all plaquettes of the initial lattice have occupation $\le \nR$. These sums are given by
\begin{subequations}
  \begin{align}
    \Y_0 &= Z_0 \,, \\
    \Y_1 &= Z_{(1)} = V Z_{(1|)} \,, \\
    \Y_2 &= \Theta(\nR-2) Z_{(2)} + Z_{(1,1)}
    = \Theta(\nR-2) V Z_{(2|)} + \frac V2 Z_{(1|1)} \,, \\
    \Y_3 &= \Theta(\nR-3) Z_{(3)} + Z_{(2,1)} + Z_{(1,1,1)}
    = \Theta(\nR-3) V Z_{(3|)} + V Z_{(2|1)} + \frac V3 Z_{(1|1,1)} 
  \end{align}
\end{subequations}
and so on for larger $n$. In
the second step of the last three lines we have used translational invariance as in \eqref{eq:Zi_Z(i)}.

\subsection{Matching the coefficients}
\label{sec:matching}

To determine the desired $Z_n$ we write them as $Z_n=a_n\X_n+b_n\Y_n$ and obtain a system of equations to be solved for the coefficients $a_n$, $b_n$, and $c_n$.
In \cref{app:Trans_inv} we show that a solution for any $\nmax$ and $\lfloor \nmax/2 \rfloor\le \nR\le \nmax-1$ is given by
\begin{equation}\label{eq:abc_k}
  c_n = \begin{cases}
    n/\nmax & \text{for } n\leq \nR\,, \\
    1 & \text{for } n> \nR\,,
  \end{cases}
  \qquad   
  b_n = 1-\frac{na_n}{V\nmax} 
  \quad\text{with} \quad
  a_n = 
  \begin{cases}
  \text{arbitrary} & \text{for } n\le \nR\,, \\
  V & \text{for } n> \nR \,.
  \end{cases}
\end{equation}
For $n=0$ we have $b_0=1$, and $a_0$ is irrelevant since $\X_0 = 0$, i.e., $Z_0 = \Y_0$.  
For $1\le n\le \nR$, choosing $a_n=V \nmax/n$ results in $ b_n=0$. If in addition we choose $\nR=\nmax-1$ we have $b_n=0$ for all $n\ge1$. Equivalent solutions can be obtained by rescaling $a_n\to ra_n $ and $c_n\to c_n/r$ for all $n$ and arbitrary $r$.
All results presented in \cref{sec:results} were obtained with $n_R=\lfloor \nmax/2 \rfloor$ and $a_n=V\nmax/n$ for $1\le n\le n_R$. We did not perform a systematic study of the dependence of the results on the choice of $n_R$ and $a_n$, because such a study would be quite expensive numerically.

\section{Results}
\label{sec:results}

In this section we present results obtained with the OS-GHOTRG method in two dimensions for $\Nc=3$, $\Nf=1,2$, and symmetric lattices of size $\Lt\times L_\text{s}$ with $\Lt = L_\text{s} =L$. We use the SuperQ truncation method of \cref{sec:truncation} unless stated otherwise.

\subsection{Observables}
\label{sec:observables}

We will compute the free-energy density
\begin{equation}\label{eq:fdef}
  f\equiv-\frac1V\ln Z
\end{equation}
(which we will often refer to as the free energy for simplicity),
the chiral condensate
\begin{equation}
  \label{eq:condensate}
  \Sigma \equiv \frac1V\frac{\partial \ln Z}{\partial m}  \,,
\end{equation}
and the quark-number density
\begin{equation}
  \rho \equiv \frac1V\frac{\partial \ln Z}{\partial \mu}  \,,
\end{equation}
where $m$ and $\mu$ stand for the mass and the chemical potential of one particular flavor.
In \cite[Sec.~7]{Samberger:2025hsr} we have discussed strategies to obtain observables from the expansion \eqref{eq:Zsce} of $Z$ in $\beta$. An obvious possibility would be to insert \eqref{eq:Zsce} in $\ln Z$ for the desired values of $\beta$. However, we concluded that more reliable results are obtained if we instead expand the free energy, and hence also the observables, directly in $\beta$ and truncate the expansion at $\nmax$.
The expansion of $-f$ reads
\begin{equation}\label{eq:logZexp}
	\frac1V\ln Z(\beta)=\sum_{n=0}^{\nmax} f_n \beta^{n}+\mathcal{O}(\beta^{\nmax+1})\,,
\end{equation}
where the minus sign is absorbed in the $f_n$. 
For example, up to order three we obtain with $\Zbar_n\equiv Z_n/Z_0$
\begin{equation}
	\frac1V\ln Z(\beta) 
	= \frac1V\ln Z_0
	+\frac{\beta}{V}\Zbar_1+\frac{\beta^2}{V}\left(\Zbar_2-\frac12\Zbar_1^2\right)
	+\frac{\beta^3}{V}\left(\Zbar_3-\Zbar_1\Zbar_2+\frac{1}{3}\Zbar_1^3\right)
	+\mathcal{O}(\beta^{4})\,.
	\label{logZexp}
\end{equation}
Taking the derivative with respect to $m$ and $\mu$ yields
\begin{subequations}\label{eq:cc_expansion}
  \begin{align}
    \Sigma &= \sum_{n=0}^{\nmax} \Sigma_n\beta^n+\mathcal{O}(\beta^{\nmax+1})\quad\text{with}\quad
    \Sigma_n\equiv\frac{\partial f_n}{\partial m}\,, \label{eq:cc_tensor}\\
    \rho &= \sum_{n=0}^{\nmax} \rho_n\beta^n+\mathcal{O}(\beta^{\nmax+1})\quad\text{with}\quad
    \rho_n\equiv\frac{\partial f_n}{\partial \mu} \,. \label{eq:rho_tensor}
  \end{align}
\end{subequations}
Due to time-reversal invariance, the partition function is invariant under $\vec\mu\to-\vec\mu$ with $\vec\mu=(\mu_1,\ldots,\mu_{\Nf})$.  Therefore the quark-number density is an odd function of $\vec\mu$, and hence $\rho_n=0$ for all $n$ at $\vec\mu=0$. In \cref{app:large_mu} we show that if $\mu\to\pm\infty$ for one particular flavor, then for this flavor we have $\Sigma_n\to0$ for all $n$, $\rho_0\to\pm\Nc$, and $\rho_n\to0$ for $n\ge1$. In \cref{app:large_mass} we show that for $\Nf=1$ and large $m$, $\Sigma$ and $\rho$ are given by the asymptotic expressions \eqref{eq:Sigma_largem} and \eqref{eq:rho_largem}, respectively. For $m\to\infty$ we then obtain $\Sigma_n\to0$ and $\rho_n\to0$ for $n < \Lt$.

For the OS-GHOTRG results presented below, the derivatives in \eqref{eq:cc_expansion} were computed numerically using a stabilized finite-difference scheme \cite{Bloch:2021mjw}.

\subsection{Applicability of the strong-coupling expansion}
\label{sec:applicability}

In this subsection we discuss the applicability of the strong-coupling expansion, both in the vicinity of and away from a phase transition. The arguments are generic, but to be specific we consider the phase transition as a function of $\mu$. 

Away from a phase transition we use \eqref{eq:cc_expansion} and expect $\Sigma_n/\Sigma_0$ and $\rho_n/\rho_0$ to be independent of $V$ for large $V$. Therefore in this case the range in $\beta$ for which \eqref{eq:cc_expansion} provides a reasonable approximation is independent of $V$. 

We now analyze the behavior of a generic observable $\Omega$ in the vicinity of a phase transition. To this end we consider a function $\omega(x)$ with a smooth crossover at $x=0$, e.g., $\omega(x)=\tanh(x)$ or $\operatorname{arctan}(x)$ or $\operatorname{erf}(x)$, and derivatives $\omega^{(n)} = \partial^n\omega/\partial x^n$. Near the phase transition we assume 
\begin{equation}
  \label{eq:generic_model}
  \Omega(\mu)\sim \omega(a(\mu-\muc))\,,
\end{equation}
where $\muc$ is the critical chemical potential and we have omitted a possible constant offset. We assume $a\sim V^\gamma$ ($\gamma>0$), which implies that the transition becomes steeper with increasing volume. For the time being we assume for simplicity that the width of the transition, which is proportional to $1/a$, is independent of~$\beta$ and that the critical chemical potential is linear in $\beta$, i.e., $\muc(\beta)=\muc_0 + \beta \muc_1$. The strong-coupling expansion of $\Omega$ at fixed $\mu$ then reads
\begin{equation}
  \label{eq:generic_observable}
  \Omega(\mu,\beta)=\sum_{n=0}^{\nmax}\Omega_n(\mu)\beta^n+\mathcal{O}(\beta^{\nmax+1})\quad\text{with}\quad
  \Omega_n(\mu) \sim \frac1{n!} (-a \muc_1)^n \omega^{(n)}(a(\mu - \muc_0)) \sim a^n\,.
\end{equation}
This implies that the range of $\beta$ for which \eqref{eq:generic_observable} gives reasonable approximations is limited by $\beta_\text{max} \sim 1/a \sim V^{-\gamma}$.

Let us now drop the assumptions on the $\beta$ dependence of $a$ and $\muc$ and assume polynomial approximations in $\beta$ with coefficients $a_n$ and $\muc_n$, respectively. We expect the $a_n/a_0$ and $\muc_n$ to have finite limits for large $V$. Therefore the polynomial approximations for $a(\beta)$ and $\muc(\beta)$ should give reasonable results in a range of $\beta$ that is independent of $V$. 
In the same range of $\beta$, we expect \eqref{eq:generic_model} to give a reasonable approximation to the observable.

The difference in the ranges of $\beta$ ($\sim1/V^\gamma$ or independent of $V$) can be understood more generally. Near the phase transition, the function $\Omega(\mu)$ is very steep (with slope $\sim a$), while the inverse function $\mu(\Omega)$ is nearly flat (with slope $\sim1/a$). As a function of $\beta$, the steep function is shifted horizontally since $\muc$ is a function of $\beta$, see \cref{fig:full_chiral_condensate_and_baryon_density} below.
At fixed $\mu$ close to $\muc$, the observable therefore changes drastically as a function of $\beta$. Therefore it is not a good idea to expand $\Omega(\mu,\beta)$ in $\beta$ at fixed $\mu$ close to $\muc$, since the coefficients of this expansion would need to be large, see \cref{fig:obs_L=32} below. In contrast, expanding $\muc$ in $\beta$ corresponds to an expansion of the inverse function $\mu(\Omega,\beta)$ in $\beta$. Since $\mu(\Omega)$ is nearly flat we do not expect large corrections.

The arguments made in this subsection will be confirmed by the numerical results for the observables $\Sigma$ and $\rho$ presented in the following. However, we will see that $a_n/a_0$ for $n\ge1$ and $\muc_n$ for $n\ge2$ become increasingly difficult to determine for increasing $V$. Since we expect $a_n/a_0$ and $\muc_n$ to remain finite for large $V$, one possible way to deal with this issue is to combine $a_0$ determined from large $V$ with $a_n/a_0$ and $\muc_n$ determined from small $V$ to still obtain reasonable approximations for the observable even for large $V$, again in a range of $\beta$ that is independent of $V$.

\subsection{Validation of OS-GHOTRG with analytical results and Monte Carlo data}

As a consistency check for the correct implementation of the OS-GHOTRG method we first consider $\Nf=1$ and compare tensor data with analytical results for $L=2$, see \cite[App.~B]{Samberger:2025hsr}, and with Monte Carlo data for $L=8$.

\begin{figure}
	\centering{
		\includegraphics[width=0.32\textwidth]{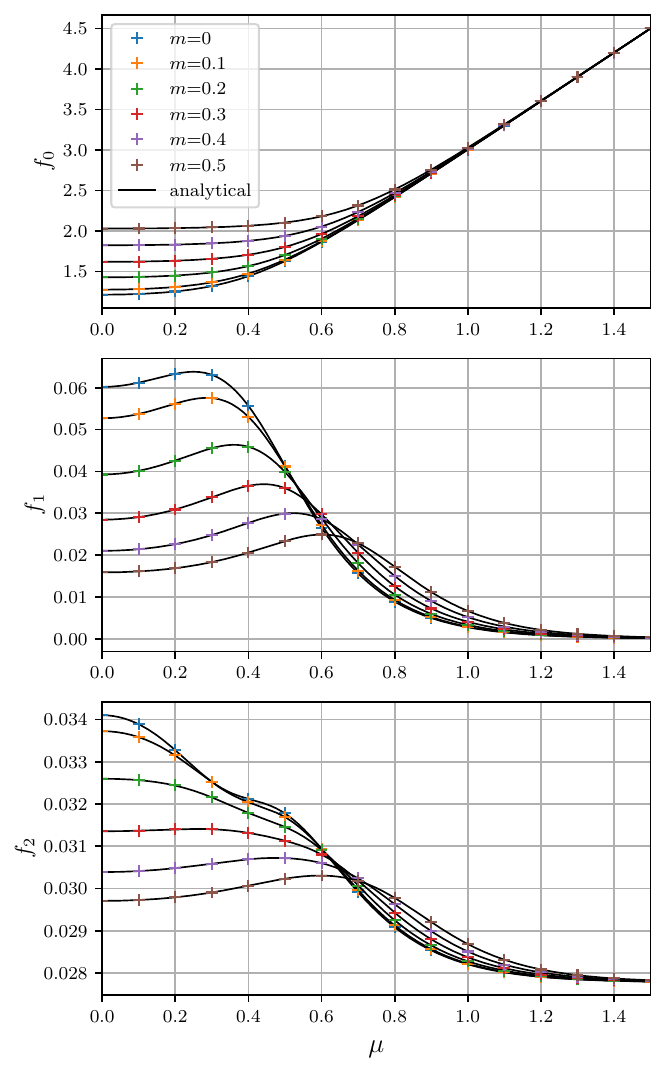}
		\includegraphics[width=0.32\textwidth]{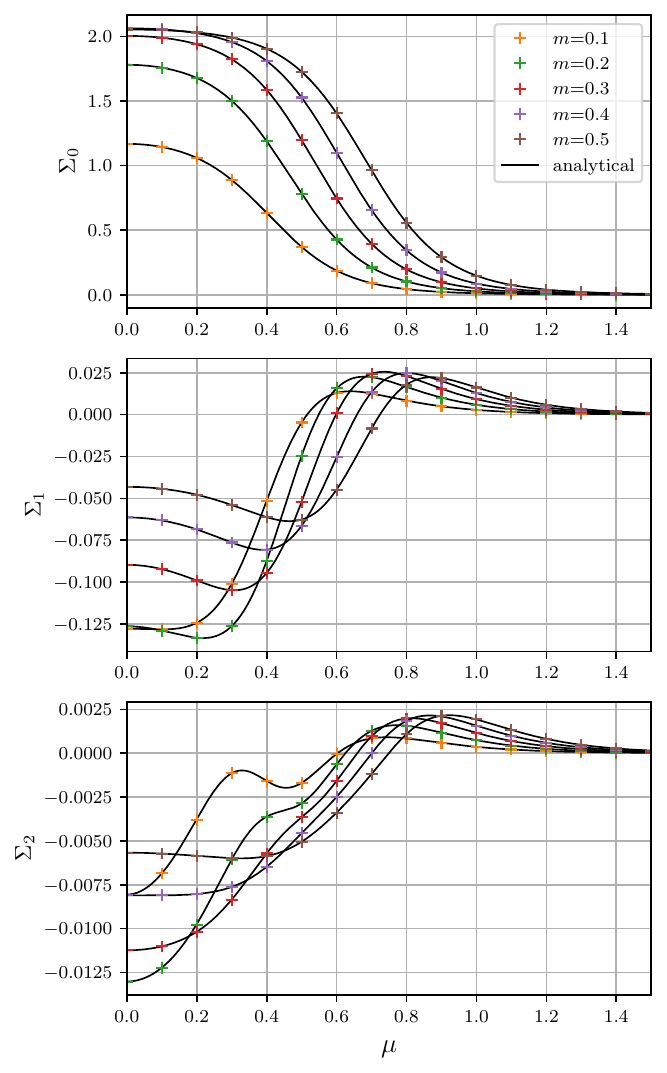}
		\includegraphics[width=0.32\textwidth]{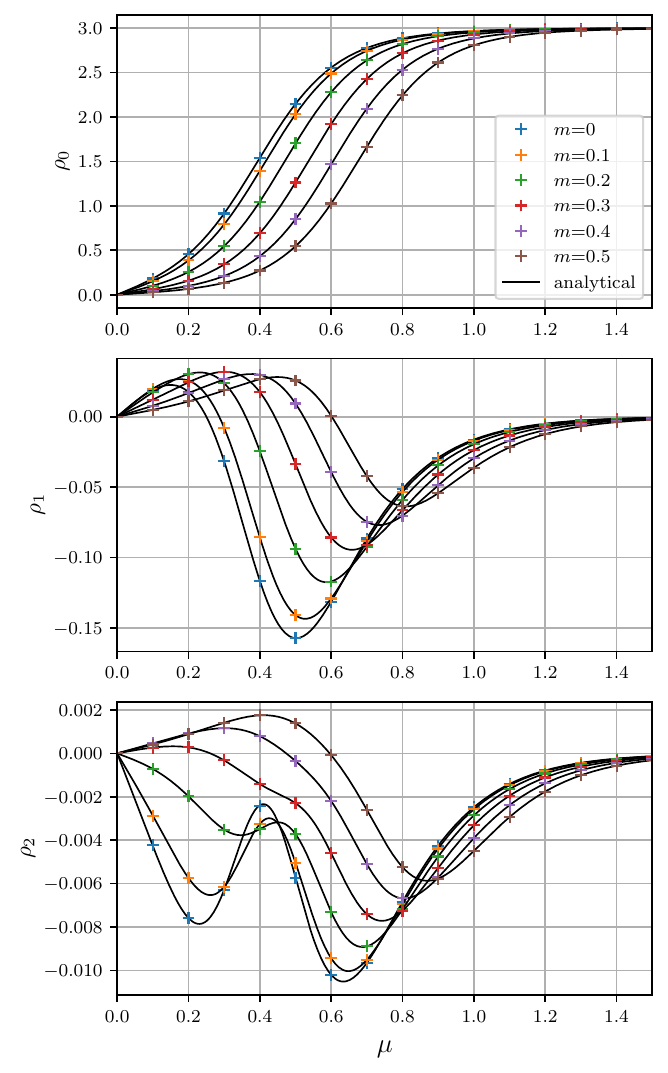}
	}
	\caption{Coefficients of the free energy (left), the chiral condensate (middle), and the quark-number density (right) for $\nmax=2$ as a function of $\mu$ for different values of $m$ on a $2\times 2$ lattice. We compare the analytical results obtained from \cite[App.~B]{Samberger:2025hsr} to the OS-GHOTRG results with bond dimension $D=230$. For $\Sigma$ we do not show the $m=0$ result, which is trivially zero on a finite lattice.\label{fig:coeff_vs_mu_2x2_OS}}
\end{figure}

In \cref{fig:coeff_vs_mu_2x2_OS} we show the coefficients $f_n$, $\Sigma_n$, and $\rho_n$ for $L=2$ as a function of the chemical potential for $\nmax=2$ and for different masses. The tensor results are obtained for bond dimension $D=230$.
The data agree nicely with the analytical results (solid lines). 
For $\nmax=3$ (not shown in the figure) larger bond dimensions are necessary to obtain agreement between the numerical and analytical results, except for large $m$ or large $\mu$.

\begin{figure}
  \centering
  \includegraphics[width=0.5\textwidth]{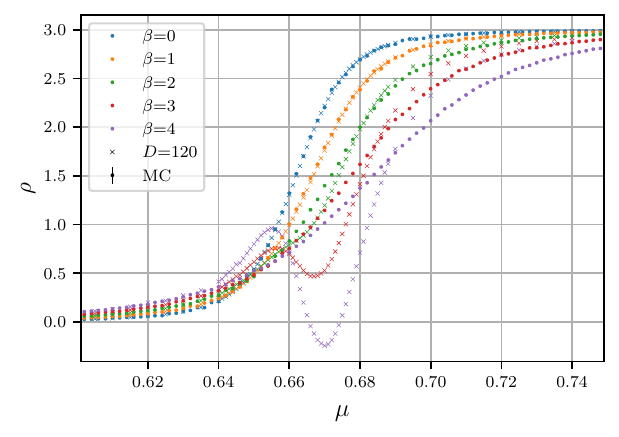}
  \caption{Quark-number density $\rho$ as a function of $\mu$ for $L=8$, $m=0.5$,  $\nmax=2$, and various values of $\beta$ obtained from Monte Carlo simulations (circles) and from the OS-GHOTRG method with bond dimension $D=120$ (crosses). The systematic deviations for larger $\beta$ are expected, see \cref{sec:applicability}.} 
\label{fig:8x8}
\end{figure}

In \cref{fig:8x8} we show the quark-number density as a function of $\mu$ near the phase transition for $L=8$ and various values of $\beta$ and compare results obtained from the OS-GHOTRG method using \eqref{eq:rho_tensor} with $\nmax=2$ and from Monte Carlo simulations using reweighting. For larger lattices the reweighted Monte Carlo method can no longer be used as the sign problem becomes unmanageable. We observe that the OS-GHOTRG results agree very well with the Monte Carlo data for small $\beta$. For larger $\beta$ we observe deviations as expected, since the validity of the expansion is limited as explained in \cref{sec:applicability}. 

\subsection{Behavior of the singular values}

\begin{figure}
	\centering{
		\includegraphics[width=0.29\textwidth]{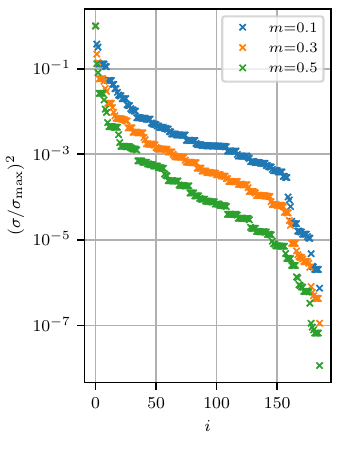}\hfill
		\includegraphics[width=0.29\textwidth]{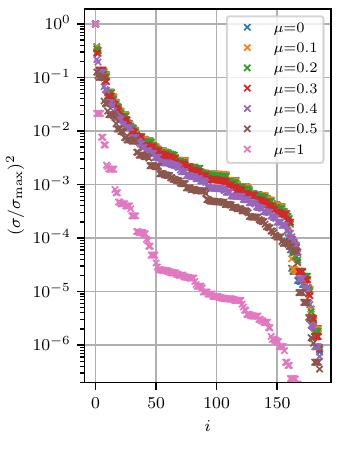}\hfill
		\includegraphics[width=0.29\textwidth]{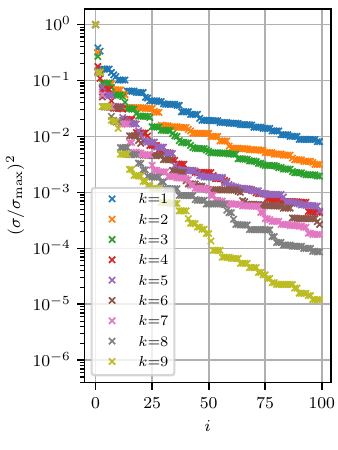}
	}
	\caption{Relative size of the squared singular values of $R\star R$, enumerated by $i$. Left: First blocking step for various values of $m$ at $\mu=0$. Middle: First blocking step for various values of $\mu$ at $m=0.1$. Right: Successive blocking steps $k$ at $m=0.1$ and $\mu=0$. All three plots are for $\nmax=2$. We observe many plateaus of nearly degenerated singular values.\label{fig:s_values}
	}
\end{figure}

We now analyze the fall-off of the squared singular values of $R\star R$, see \eqref{eq:Xblocking}, in different blocking steps for $\nmax=2$ and various values of $m$ and $\mu$, still for $\Nf=1$. In \cref{fig:s_values} (left) we show all singular values obtained in the first blocking step for $\mu=0$ and different values of $m$. We observe that the singular values decrease faster for larger mass.
Similarly, in \cref{fig:s_values} (middle) we show the singular values for $m=0.1$ and different values of $\mu$. 
The singular values decrease faster for larger $\mu$. 
We do not observe a change in the behavior of the singular values close to the phase transition, which is located at about $\muc=0.4$ for $L=2$ and $m=0.1$, as can be seen in \cref{fig:coeff_vs_mu_2x2_OS}.
Finally, \cref{fig:s_values} (right) shows the singular values for nine successive blocking steps for $m=0.1$, $\mu=0$, and $D=100$. We observe that the singular values decrease faster in later blocking steps. 

\subsection{Results for larger lattices near the phase transition}
\label{sec:qnd}

\begin{figure}
	\centering
	{
		\hfill
		\includegraphics[width=0.45\textwidth]{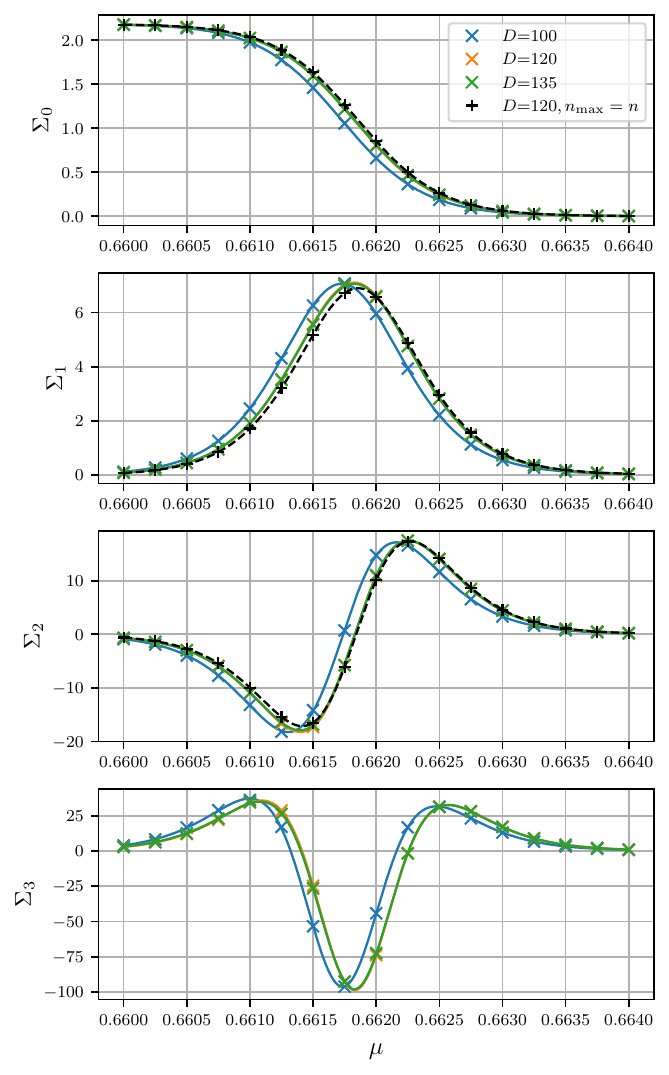}\hfill
		\includegraphics[width=0.45\textwidth]{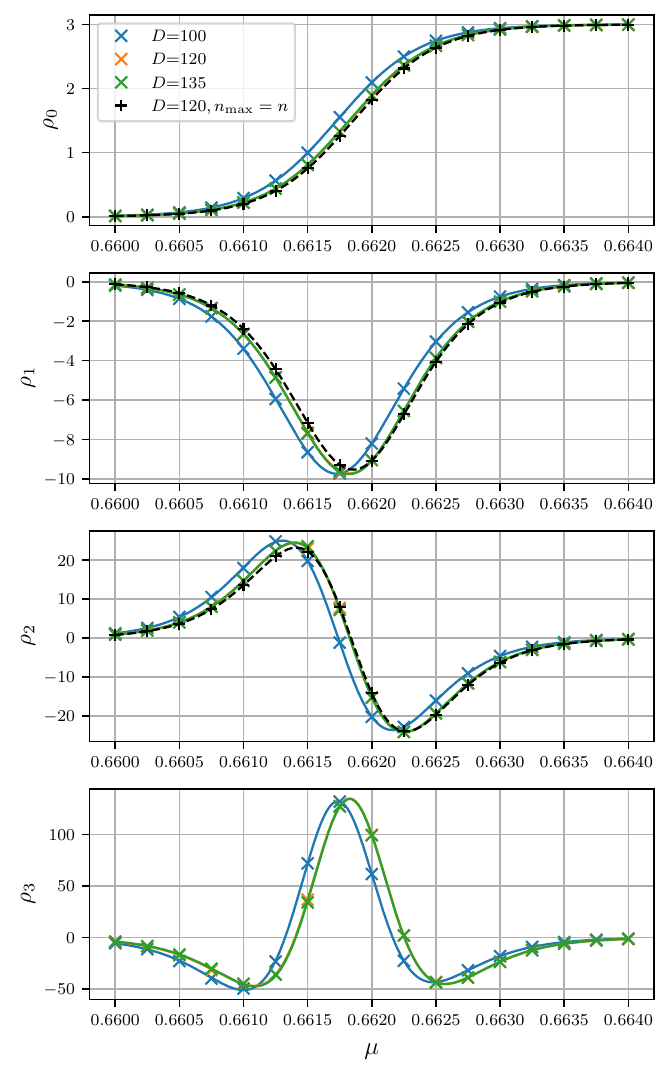}
		\hfill
	}
	\caption{Coefficients of the chiral condensate (left) and the quark-number density (right) as a function of $\mu$ near the phase transition for $\nmax=3$ with $L=32$ and $m=0.5$ using the Xie et al. truncation method for various bond dimensions $D$. We observe excellent agreement between the tensor data (crosses) and the model functions (solid lines) given in \cref{app:Coeff_exp}. The large-$\mu$ limits of the coefficients $\Sigma_n$ and $\rho_n$, see \cref{sec:observables}, are consistent with the analytical results. In the bottom row, the data points for $\nmax=n$ are already given by the orange crosses.}
	\label{fig:obs_L=32}
\end{figure}

\begin{figure}
  \centering
  \includegraphics[width=0.49\textwidth]{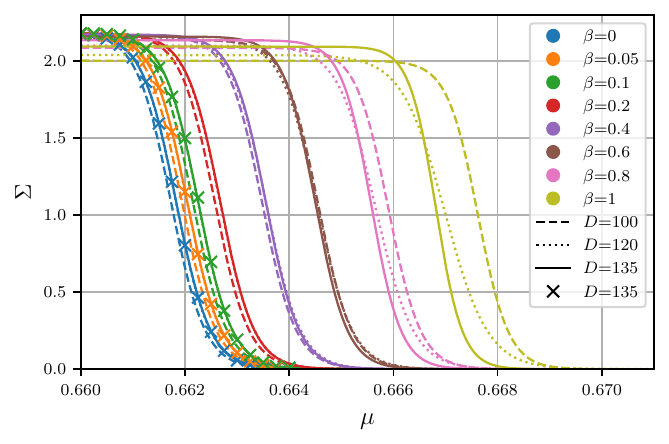}\hfill
  \includegraphics[width=0.49\textwidth]{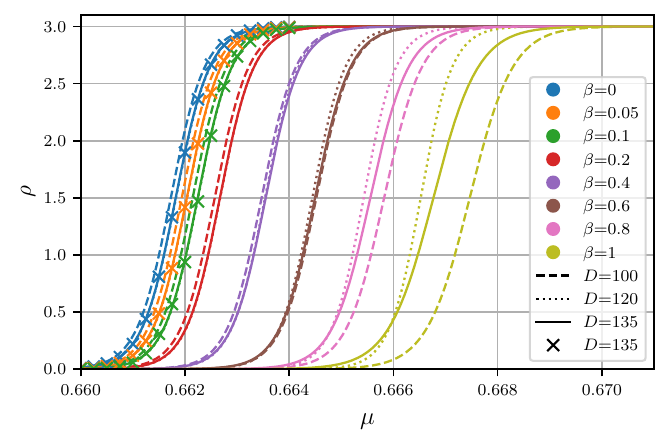}
  \caption{Chiral condensate (left) and quark-number density (right) as a function of $\mu$ for $m=0.5$, $L=32$, different bond dimensions, and several values of $\beta$ for $\nmax=3$. The data points (crosses) were obtained from \eqref{eq:cc_expansion}, while the lines correspond to the tanh model of \eqref{ccfit} and \eqref{eq:fit-ansatz}, respectively. In both plots, we observe an increasing critical chemical potential with increasing coupling parameter $\beta$.}
  \label{fig:full_chiral_condensate_and_baryon_density}
\end{figure}

\begin{figure}
	\centering
	\includegraphics[width=0.9\textwidth]{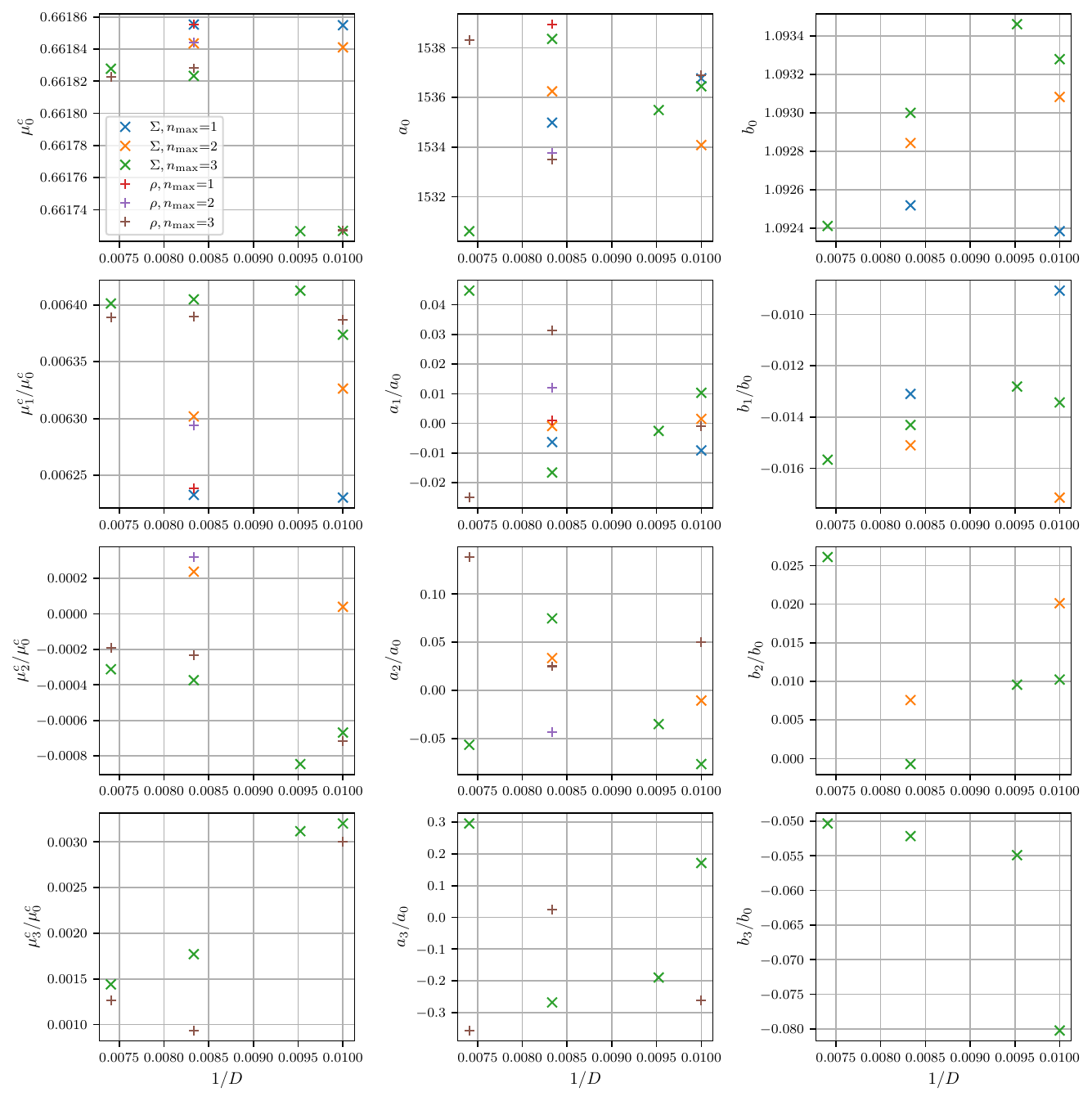}
	\caption{Dependence of the coefficients of the tanh model, see \eqref{eq:model_sigma} and \eqref{expansions}, on the bond dimension $D$ for $\nmax=1,2,3$, $m=0.5$, and $L=32$. To remove the overall scale we plot the coefficients for $n\ge1$ relative to the coefficient for $n=0$.}
	\label{fig:coeff_vs_D}
\end{figure}

\begin{figure}
	\centering
	\includegraphics[width=12cm]{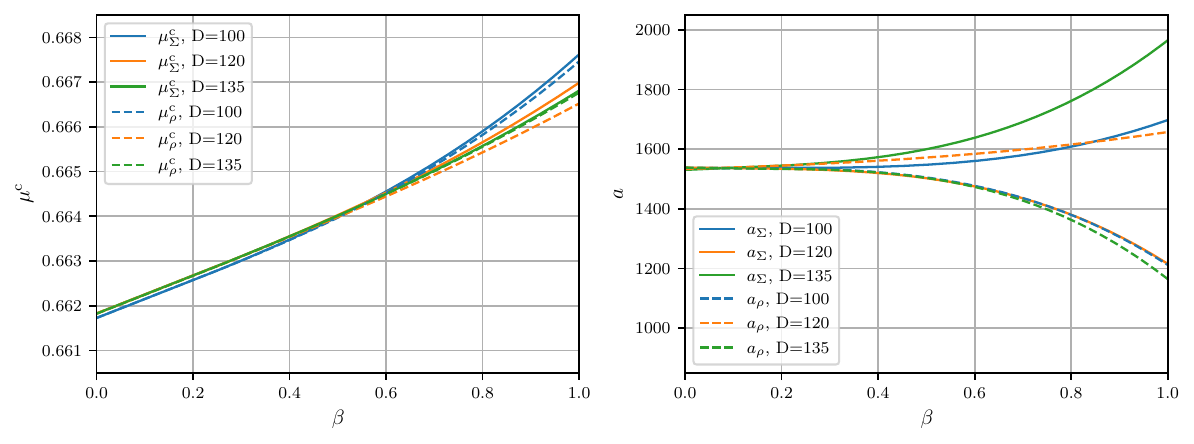}
	\caption{Critical chemical potential and transition sharpness as a function of $\beta$ for the chiral condensate and the quark-number density for $\nmax=3$, $m=0.5$, and $L=32$.}
	\label{fig:coeff}
\end{figure}

\begin{figure}
	\centering
	\includegraphics[width=0.9\textwidth]{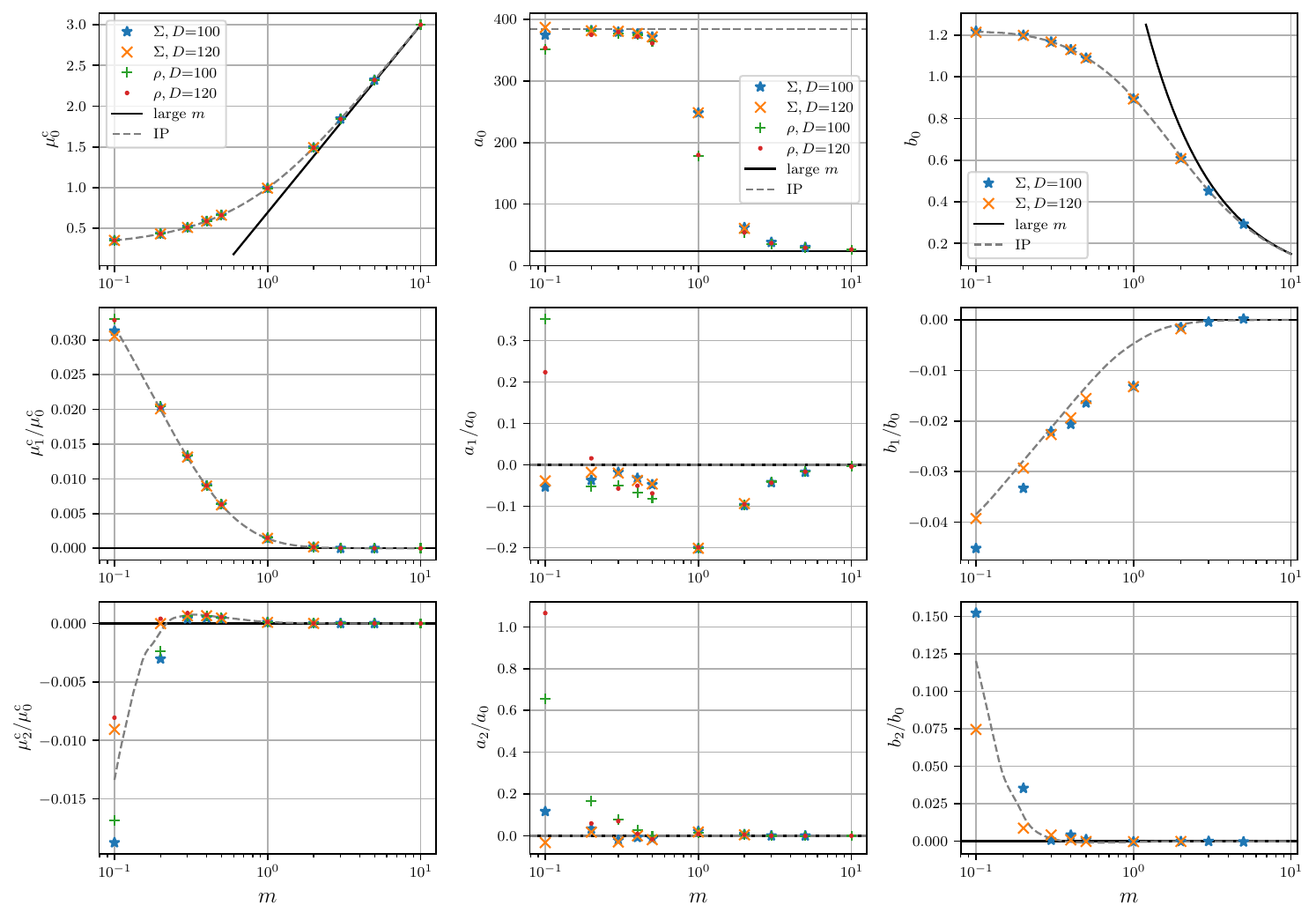}
	\caption{Fitted coefficients of the tanh models as functions of the quark mass $m$. In contrast to the previous figures, we set $L=16$ in order to be able to reach small values of $m$ with manageable numerical effort (determined by $D$). Smaller $m$ requires higher bond dimensions to obtain convergence, see \cref{fig:s_values}. In the OS-GHOTRG method, larger $L$ also requires higher bond dimensions due to volume-dependent cancellations in \eqref{logZexp} for $n\ge2$, see \cite[Sec.~7]{Samberger:2025hsr}. The large-$m$ and interpolation (IP) results are given in \cref{app:large_mass} and \cref{app:interpolation}, respectively.}
	\label{fig:coeff_vs_m}
\end{figure}

\begin{figure}
	\centering
	\includegraphics[width=0.7\textwidth]{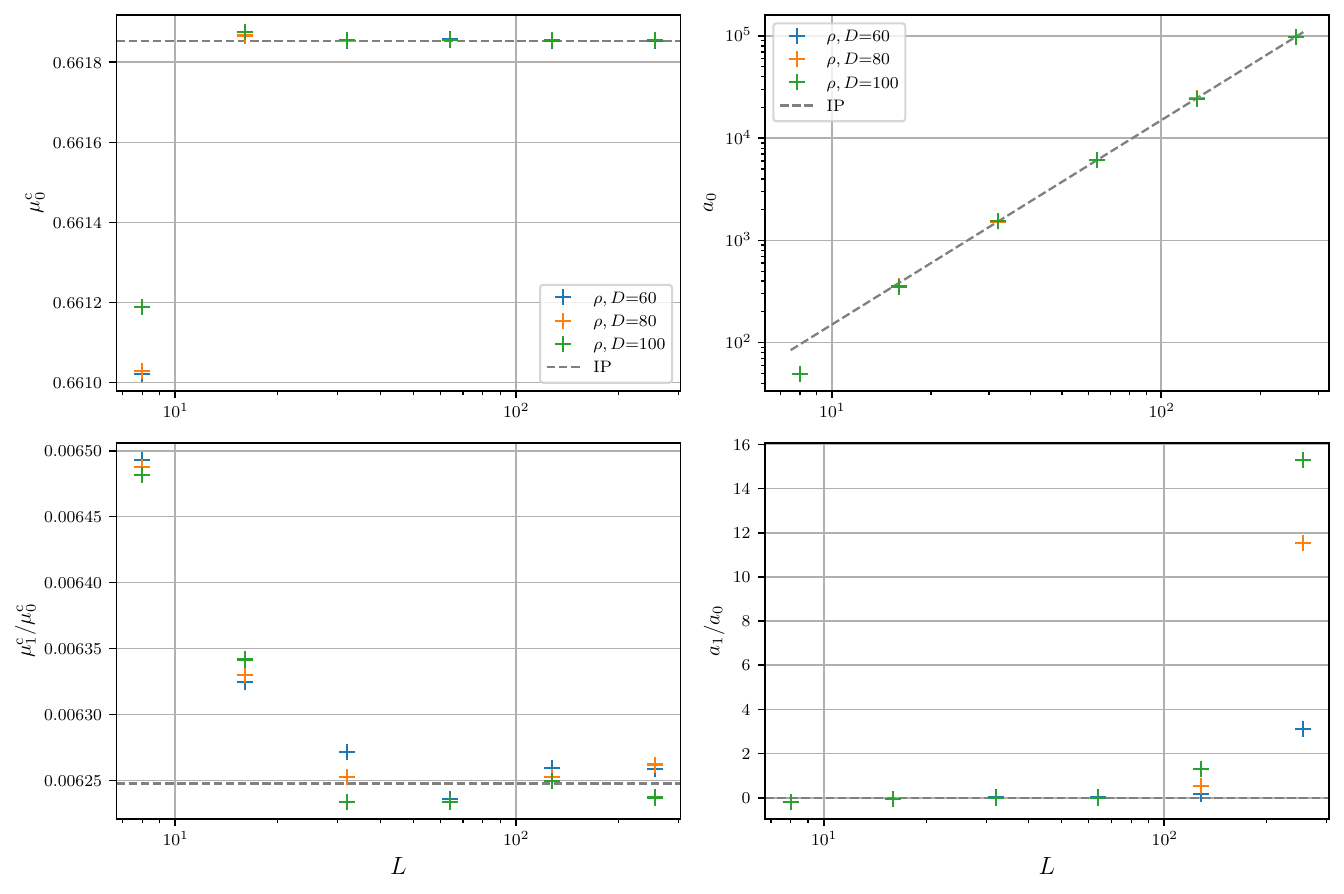}
	\caption{Fitted coefficients of \eqref{expansions} as functions of $L$ for $\nmax=1$ and $m=0.5$. Note that $\muc_0$ and $\muc_1$ vary only very slightly with $L$. The interpolation (IP) results are given in \cref{app:interpolation}.}
	\label{fig:coeff_vs_V}
\end{figure}

In this subsection we present results for larger lattices, still for $\Nf=1$, and introduce a fit ansatz based on \cref{sec:applicability} that describes the data very nicely in the vicinity of the phase transition. In \cref{fig:obs_L=32,fig:full_chiral_condensate_and_baryon_density,fig:coeff_vs_D,fig:coeff} we present OS-GHOTRG results obtained from a particular example, i.e., $L=32$, $m=0.5$, $\nmax=3$, and several bond dimensions $D$. In \cref{fig:coeff_vs_m,fig:coeff_vs_V} we then analyze the dependence of the OS-GHOTRG results on $m$ and $L$.

Figure~\ref{fig:obs_L=32} shows the coefficients of the chiral condensate and the quark-number density as a function of $\mu$ near the phase transition for three different bond dimensions. 
The small discrepancies between the data for different values of $D$ imply that the results have not yet fully converged. As a consistency check, we also show the coefficients $\Sigma_n$ and $\rho_n$ obtained with $\nmax=n$, which converge for smaller values of $D$.

\afterpage{\clearpage}

In order to model the behavior of these coefficients in the vicinity of the phase transition we introduce a fit ansatz for each observable, motivated by Fermi-Dirac statistics and the representation of $Z$ in \eqref{eq:Z_cosh_Nf1}.\footnote{This ansatz will be substantiated further in \cref{app:large_mass,app:interpolation}.} For the chiral condensate we use\footnote{An improved fit ansatz would be $\Sigma^{\text{model}}(\mu,\beta)+\Sigma^{\text{model}}(-\mu,\beta) - 2b_\Sigma(\beta)$, which is even in $\mu$ as required.  However, for $\mu \geq 0$ the additional term is negligible if $a_\Sigma\mu_\Sigma^\text{c}\gg1$, which is the case for the quark masses and lattice sizes we use. \label{footnote:cc_model}}
\begin{equation}\label{ccfit}
  \Sigma^{\text{model}}(\mu,\beta)=b_\Sigma(\beta)\left(1-\tanh\left[a_\Sigma(\beta)\left(\mu-\mu^{\text{c}}_\Sigma(\beta)\right)\right]\right)\\
\end{equation}
with parameters that also depend on $m$ and the lattice size.
The parameters $\mu^{\text{c}}_{\Sigma}$ and $a_\Sigma$ are referred to as critical chemical potential and transition sharpness, respectively, corresponding to the chiral condensate.
Using polynomial approximations
\begin{equation}\label{eq:model_sigma}
  \mu^{\text{c}}_{\Sigma}(\beta) = \sum_{n=0}^{\nmax}\mu^{\text{c}}_{\Sigma,n}\beta^n
  \,,\quad
  a_\Sigma(\beta) = \sum_{n=0}^{\nmax}a_{\Sigma,n}\beta^n
  \,, \quad
  b_\Sigma(\beta) = \sum_{n=0}^{\nmax}b_{\Sigma,n}\beta^n
\end{equation}
for the parameters and expanding the hyperbolic tangent in $\beta$, we obtain model functions for all coefficients $\Sigma_n$, which are given in \cref{app:Coeff_exp} up to $n=3$. For every bond dimension shown in \cref{fig:obs_L=32} (left), the tensor data for $\Sigma_n$ are fitted with the model functions $\Sigma_n(\mu)$ simultaneously for all $n$. To take into account the different ranges of the $\Sigma_n$, given by $r_n=\max_\mu\Sigma_n(\mu)-\min_\mu\Sigma_n(\mu)$, we assign a weight $1/r_n^2$ to every data point.

For the quark-number density we use the fit ansatz\footnote{Again, an improved fit ansatz would be $\rho^{\text{model}}(\mu,\beta) - \rho^{\text{model}}(-\mu,\beta)$, which is odd in $\mu$.\label{footnote:rho_model}}
\begin{equation}\label{eq:fit-ansatz}
  \rho^{\text{model}}(\mu,\beta)=\frac{\Nc}2\bigl(1+\tanh\bigl[a_\rho(\beta)\bigl(\mu-\mu^{\text{c}}_\rho(\beta)\bigr)\bigr]\bigr)\,,
\end{equation}
which contains the critical chemical potential and transition sharpness corresponding to $\rho$. Both parameters also depend on $m$ and the lattice size. Using polynomial approximations
\begin{equation}\label{expansions}
	\mu^{\text{c}}_\rho(\beta)=\sum\limits_{n=0}^{\nmax}\mu^{\text{c}}_{\rho,n}\beta^n
	\quad\text{and}\quad
	a_\rho(\beta)=\sum\limits_{n=0}^{\nmax}a_{\rho,n}\beta^n
\end{equation}
and expanding the hyperbolic tangent, we obtain model functions for all coefficients $\rho_n$, also given in \cref{app:Coeff_exp} up to $n=3$. For every bond dimension, we again perform simultaneous fits of the model functions to the tensor data with weights assigned in the same way as for the chiral condensate. As shown in Fig.~\ref{fig:obs_L=32}, we find excellent agreement between the model functions and the tensor data for both observables, with 12 and 8 fit parameters for $\Sigma$ and $\rho$, respectively. Furthermore, we observe that the coefficients for $n\ge1$ become large near the phase transition, as anticipated in \cref{sec:applicability}.

In Fig.~\ref{fig:full_chiral_condensate_and_baryon_density} we plot the results for $\Sigma(\mu,\beta)$ and $\rho(\mu,\beta)$ as a function of $\mu$, close to the phase transition, for $\nmax=3$, $m=0.5$, $L=32$, and several values of $\beta$.
For small values of $\beta$ (e.g., $\beta\leq0.1$ in \cref{fig:full_chiral_condensate_and_baryon_density}) the expansions in \eqref{eq:cc_expansion} coincide with the tanh model of \eqref{ccfit} and \eqref{eq:fit-ansatz}, respectively. However, for larger values of $\beta$, we observe that the expansions in \eqref{eq:cc_expansion} exhibit increasingly unphysical behavior (not shown in the figure). This is to be expected since the coefficients $\Sigma_n$ and $\rho_n$ must become large (for $n\ge1$) near the critical chemical potential as explained in \cref{sec:applicability}, see also \cref{fig:obs_L=32}.
In contrast, in the tanh model of \eqref{ccfit} and \eqref{eq:fit-ansatz} with the expansions \eqref{eq:model_sigma} and \eqref{expansions}, the coefficients $\mu^\text{c}_n$, $a_n/a_0$, and $b_n$ do not have to become large, which is also what we find, see \cref{fig:coeff_vs_D}. 
Therefore we expect the tanh model to yield reliable results also for larger values of $\beta$ (e.g., for $\beta>0.1$ in \cref{fig:full_chiral_condensate_and_baryon_density}). 
Note that, as expected, the region of $\beta$ where the expansion \eqref{eq:cc_expansion} works is much smaller for $L=32$ than for $L=8$, see \cref{sec:applicability}. Moreover, we also note a qualitative change in the phase transition. For $L=8$ the transition sharpness decreases with $\beta$, whereas for $L=32$ it is rather insensitive to $\beta$, and it is merely the position $\muc$ of the phase transition that is shifted to larger values when $\beta$ increases. The convergence (or lack thereof) of the fit parameters as a function of the bond dimension is shown in \cref{fig:coeff_vs_D}.

In Fig.~\ref{fig:coeff} we show the $\beta$ dependence of critical chemical potential $\mu^\text{c}(\beta)$ and transition sharpness $a(\beta)$ for both observables, obtained from \eqref{eq:model_sigma} and \eqref{expansions}, using $\nmax=3$, $L=32$, and the coefficients shown in Fig.~\ref{fig:coeff_vs_D} for bond dimensions $D=100,120,135$. We observe that the results obtained from $\rho$ and $\Sigma$ coincide within the systematic errors of the tensor-network method. 


We now turn to the dependence of the OS-GHOTRG results on $m$ and $L$. In \cref{fig:coeff_vs_m} we show the coefficients of \eqref{eq:model_sigma} and \eqref{expansions} as a function of $m$. Again we observe that for both observables the critical chemical potential is very similar, and so is the transition sharpness. As expected, the critical chemical potential increases with the quark mass. Also shown in the figure are asymptotic limits for large $m$ (see \cref{app:large_mass}), which corresponds to the quenched theory. Finally, the dashed lines (marked IP) correspond to formulas obtained in \cref{app:interpolation} from interpolating $Z$ between $\mu=0$ and $\mu=\infty$. The value of $a_0=3V/2$ observed for small $m$ (for both observables) is consistent with this interpolation, see \eqref{eq:a_ip}.

In \cref{fig:coeff_vs_V} we show the dependence of the coefficients of \eqref{expansions} on $L$ for $m=0.5$. We chose $\nmax=1$ because for larger volumes larger bond dimensions are required to determine $a_n/a_0$ (for $n\ge1$) and $\muc_n$ (for $n\ge2$) reliably. This is expected in the vicinity of the phase transition, whose sharpness increases with the lattice size (with exponent $\gamma$ depending on $m$, see \cref{fig:coeff_vs_m}), see \cref{sec:applicability} and \cref{app:Coeff_exp}.  This issue already occurs for the largest values of $L$ in \cref{fig:coeff_vs_V} and becomes more severe for even larger $L$.
Nevertheless, within the accuracy of the OS-GHOTRG results, $a_1/a_0$ is independent of $L$ for large $L$, as assumed earlier. 
The coefficients for $\muc$ are nearly constant in $L$ so that the location $\muc$ of the phase transition seems to converge rapidly to its infinite-volume limit. 

\subsection{Two flavors}
\label{sec:twoflavors}

We now present results for the case of two flavors. Note that for $L=2$ the partition function and its derivatives can be computed analytically \cite{Samberger2026}. We will use these analytical results to validate the numerical results. We will also compare to Monte Carlo data obtained at $\mu=0$.

\begin{figure}
	\centering{
		\includegraphics[height=0.17\textheight]{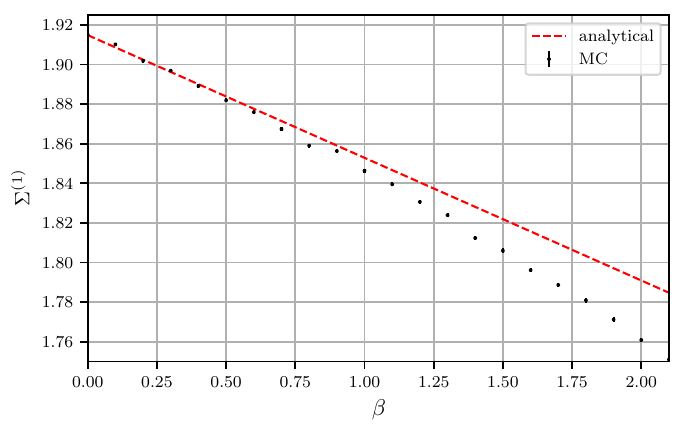}
		\includegraphics[height=0.17\textheight]{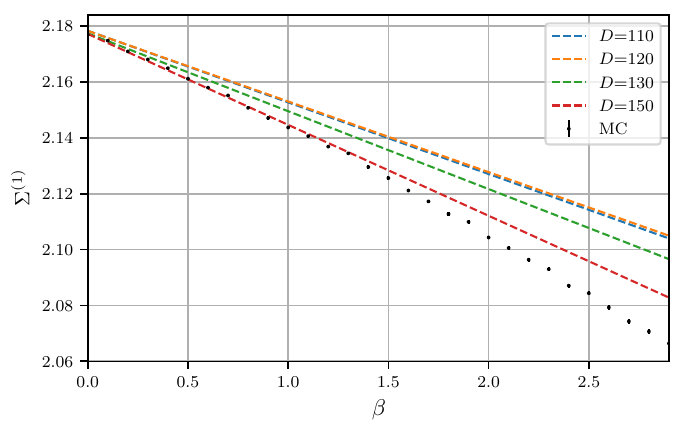}
	}
	\caption{Chiral condensate $\Sigma^{(1)}=-\partial f/\partial m_1$ for $\Nf=2$ as a function of $\beta$ for $m_1=m_2=0.5$ with $\mu_1=\mu_2=0$. Left: Comparison of analytical results obtained from the initial tensor for $\nmax=1$ with Monte Carlo data on a $2\times 2$ lattice.
	Right: Comparison of the OS-GHOTRG results for $\nmax=1$ with Monte Carlo data on an $8\times 8$ lattice.}
	\label{fig:cc1}
\end{figure}

For $\Nf=2$ the chiral condensate for particle $i$ is given by $\Sigma^{(i)}=-\partial f/\partial m_i$. In Fig.~\ref{fig:cc1} (left) we show the chiral condensate as a function of $\beta$ for $L=2$, $m_1=m_2=0.5$, and $\mu_1=\mu_2=0$ and compare analytical results \cite{Samberger2026} obtained from the initial tensor for $\nmax=1$ with Monte Carlo data. Both the infinite-coupling limit and the slope are in good agreement. In Fig.~\ref{fig:cc1} (right) we show a similar plot for $L=8$, where we compare OS-GHOTRG results obtained from \eqref{eq:cc_tensor} with various bond dimensions to Monte Carlo data. For the largest bond dimensions the slope of the tensor data  agrees well with that of the Monte Carlo calculation.

\begin{figure}
	\centering{
		\includegraphics[width=0.32\textwidth]{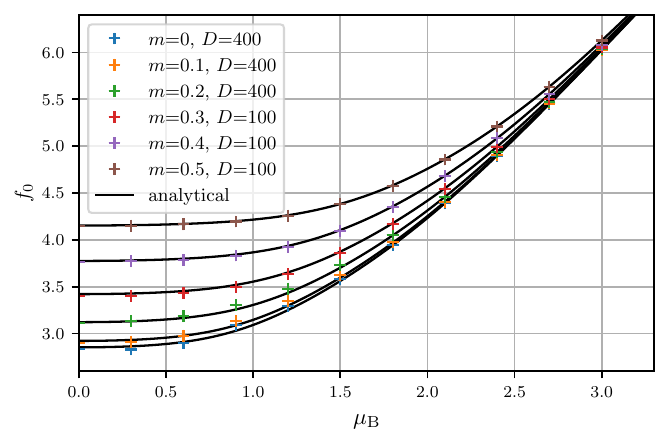}
		\includegraphics[width=0.32\textwidth]{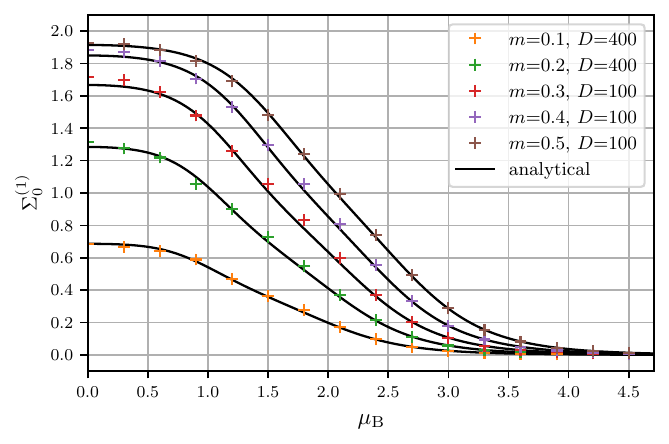}
		\includegraphics[width=0.32\textwidth]{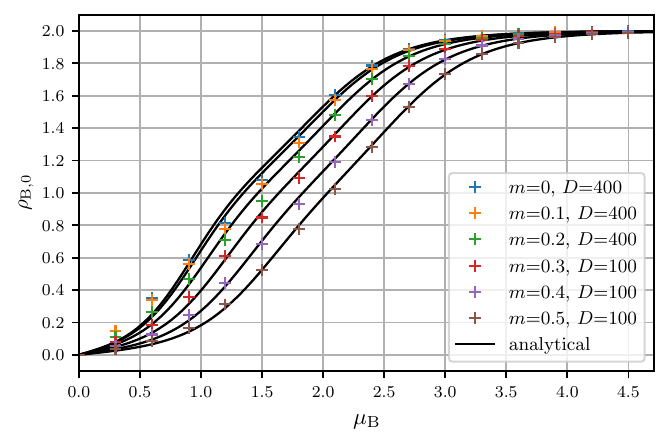}
	}
	\caption{Plot of $-f=(\ln Z)/V$ (left), $\Sigma^{(1)}$ (middle), and $\rho_B$ (right) as a function of $\mub$ for $\Nf=2$, $\mui=0$, $m_1=m_2=m$, and $\nmax=0$, i.e., in the infinite-coupling limit, on a $2\times 2$ lattice. The tensor-network results are compared with analytical tensor results.
	\label{fig:f_2x2_Nf2}}
\end{figure}
\begin{figure}
	\centering{
    \hfill
	\includegraphics[width=0.32\textwidth]{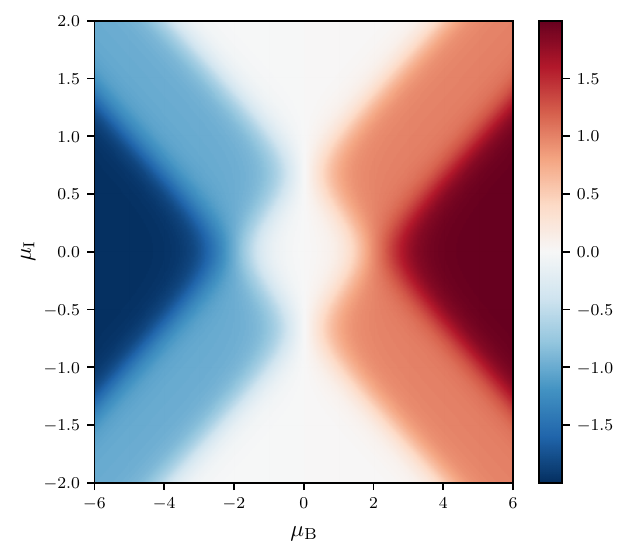}
    \hfill
	\includegraphics[width=0.32\textwidth]{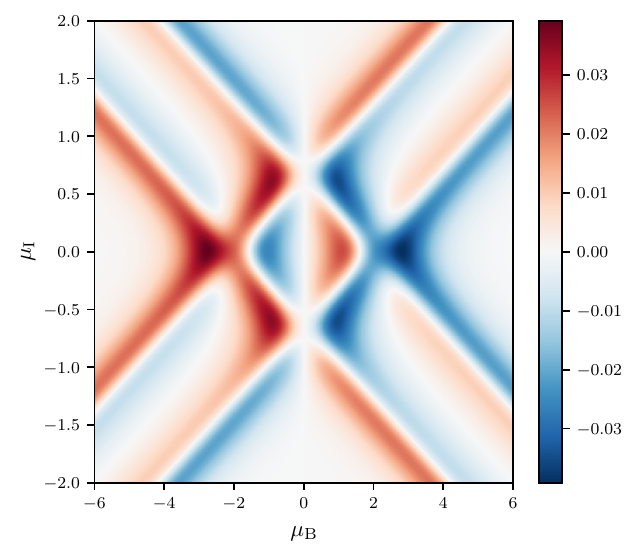}
    \hfill}
	\caption{Analytical results for $\rho_{\text{B},0}$ (left) and $\rho_{\text{B,1}}$ (right) as a function of $\mub$ and $\mui$ for $L=2$, $\nmax=1$, and $m_1=m_2=0.5$.}
	\label{fig:contour}
\end{figure}
\begin{figure}
	\centering
	\includegraphics[width=60mm]{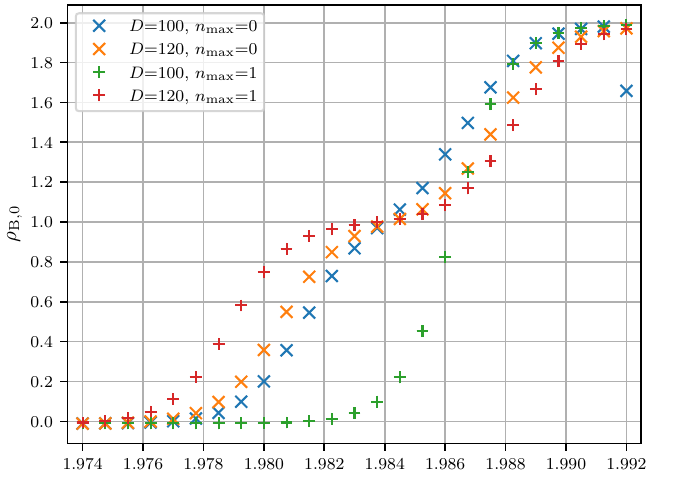}
	\hspace{3mm}
	\includegraphics[width=60mm]{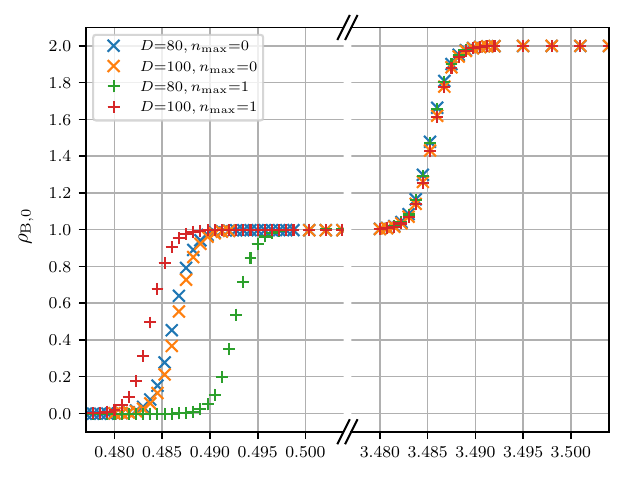}
	\\
	\includegraphics[width=60mm]{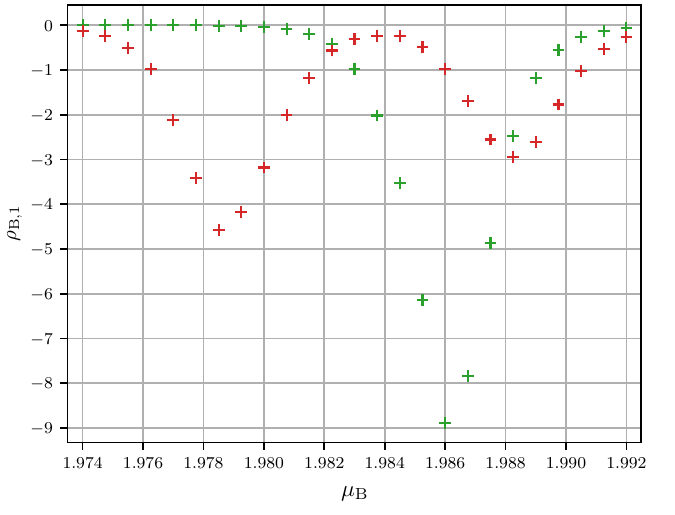}
	\hspace{3mm}
	\includegraphics[width=60mm]{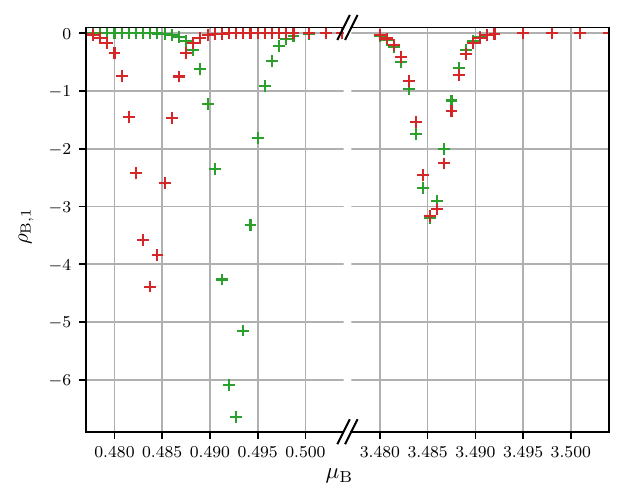}
	\caption{$\rho_{\text{B},0}$ (top) and $\rho_{\text{B},1}$ (bottom) versus $\mub$ for $\mui=0$ (left) and $\mui=0.5$ (right), $\Nf=2$, $m_1=m_2=0.5$, $L=32$, and using the Xie et al. truncation method.\label{fig:fi_vs_mub_L=32_m=0.5_mui=0.5_o=1}}
\end{figure}

In \cref{fig:f_2x2_Nf2} we turn on the chemical potentials and show the infinite-coupling limit, i.e., $\nmax=0$, of the free energy $f$, the chiral condensate $\Sigma^{(1)}$, and the baryon-number density $\rho_B=-\partial f/\partial \mu_B$ as a function of the baryon chemical potential $\mub=3(\mu_1+\mu_2)/2$ for various mass values and zero isospin chemical potential $\mui=(\mu_1-\mu_2)/2$, still for $L=2$. 
The tensor results again agree very well with the analytical predictions \cite{Samberger2026}. 
Note that the smaller masses required a larger bond dimension to obtain an acceptable accuracy. This is related to the behavior of the singular values, see also Fig.~\ref{fig:s_values}.

Next, we turn on the isospin chemical potential, i.e., we set $\mu_1\neq\mu_2$. In \cref{fig:contour} we show analytical results, obtained with the $\nmax=1$ initial tensor, for the coefficients $\rho_{\text{B},0}$ and $\rho_{\text{B,1}}$ of the baryon density as a function of the baryon and isospin chemical potential for $L=2$ and $m_1=m_2=0.5$. In the heatmap of $\rho_{\text{B},0}$ we clearly recognize the pattern of the phase transition with plateaus at $\rhob=1$ and $2$, to which $\rho_{\text{B},1}$ yields small corrections as a function of $\beta$.

We now also consider larger lattices for which the OS-GHOTRG was developed. In \cref{fig:fi_vs_mub_L=32_m=0.5_mui=0.5_o=1} we show $\rho_{\text{B},0}$ and $\rho_{\text{B,1}}$ as a function of $\mub$ for $\mui=0$ (left) and $\mui=0.5$ (right) for $L=32$.
When turning on the isospin chemical potential we see that there are two phase transitions at different values of $\mub$, which correspond to the excitation of the quarks with different chemical potentials $\mu_1\neq\mu_2$. We observe that for the smaller values of $\mub$, the results obtained for $\nmax=1$ have not yet converged for the bond dimensions that we could reach on our computers.

\section{Conclusions}
\label{sec:conclusions}

We have presented the order-separated Grassmann higher-order tensor renormalization group (OS-GHOTRG) method to investigate lattice QCD in the strong-coupling expansion. This modification of the GHOTRG method enables us to compute the expansion coefficients of the partition function, the free energy, and thermodynamical observables order by order in the strong-coupling expansion up to any order, for any number of dimensions, number of colors, and number of staggered fermion flavors starting from the tensor formulation derived in \cite{Samberger:2025hsr}. The crucial ingredient of the method is that at every blocking step, the entries of the coarse-grid tensor are given by a power series in $\beta$.
If the initial tensor is exact up to order $\beta^\nmax$ and contains no higher-order contributions, then the partition function computed with the OS-GHOTRG method is also exact up to this order, up to HOSVD truncation errors, and does not contain any higher-order contributions. The free energy, and hence also the thermodynamical observables, are computed up to $\beta^\nmax$ by a truncated Taylor expansion of $(\ln Z)/V$. For the reliability of the results it is crucial that this expansion contains no incomplete higher-order contributions, as these would produce errors that grow with the volume. As shown in \cite[Sec.~7]{Samberger:2025hsr}, this explains the essential improvement achieved by OS-GHOTRG compared to GHOTRG. The performance of the OS-GHOTRG method is substantially improved by making use of the translational invariance of the system.

We have validated the OS-GHOTRG method up to $\nmax=2$ on a $2\times2$ lattice by comparing tensor results order by order with analytical calculations for $\Nf=1$, and on an $8\times8$ lattice by comparing tensor results for the quark-number density with Monte Carlo data. For larger lattices ($L>8$) and with $\nmax=3$ we argued, and confirmed numerically, that if we expand the observables near the phase transition in $\beta$, the higher-order coefficients become large and the expansion has a convergence radius that decreases with the lattice volume. However, we found that the chiral condensate and the quark-number density are fitted very well by $\tanh$-based models, where the fit parameters are the critical chemical potential $\muc$, the sharpness $a$ of the phase transition, and the magnitude $b$ of the chiral condensate.
Since the fit parameters are expected to vary smoothly with $\beta$ we expressed them as polynomials in $\beta$.  The fits then allow for an extrapolation of the observables to much larger $\beta$ values, and the range of validity appears to be nearly independent of the lattice volume. We also discussed how the fit parameters depend on the bond dimension, the quark mass, and the lattice volume.
Note that we presented results for moderate values of the mass, as the computational effort rises very quickly when the mass is taken to be small.

We also presented $\Nf=2$ results, only up to $\nmax=1$ as larger bond dimensions are required to obtain reliable results. We compared observables computed with  the OS-GHOTRG method with analytical results for a $2\times2$ lattice and with Monte Carlo data on an $8\times8$ lattice and found good agreement up to higher-order corrections in $\beta$. We also showed the evolution of the observables, in particular the baryon-number density, as a function of the baryon and isospin chemical potentials for lattices up to $L=32$. 
 
In future work, it would be interesting to apply the OS-GHOTRG method in three and four dimensions, where the computational complexity and the memory requirements become much larger. It would therefore be very useful to again employ translational invariance, for which the ideas of \cref{sec:trans_inv} would need to be generalized to more than two dimensions.

\appendix

\section{Sign factor for tensor contraction}
\label{app:sign}

The sign factor in \eqref{eq:just_contraction} has been computed in \cite{Bloch:2022vqz} and \cite{Milde2023} in two and arbitrary dimensions, respectively. Assuming that the contraction direction is $\mu$, it can be written as
\begin{equation}\label{eq:sign}
  \sigma_{\f_x,\f_y} = (-1)^{f_x+f_y+g+f_X} 
\end{equation}
with 
\begin{equation}
  f_x = \f_{x,\mu}\sum_{\nu=1}^{\mu-1} \f_{x,\nu} + \f_{x,-\mu}\sum_{\nu=1}^{\mu-1}\f_{x,-\nu}
\label{hatsigma}
\end{equation}
and
\begin{equation}
  g = \Biggl(\sum_{\nu=1\atop\nu\neq\mu}^d\f_{y,-\nu}\Biggr)\Biggl(\sum_{\nu=1\atop\nu\neq\mu}^d\f_{x,\nu}\Biggr) 
+ \sum_{\rho=1\atop\rho\neq\mu}^d \f_{x,\rho} \sum_{\nu=\rho\atop\nu\neq\mu}^d \f_{y,\nu}
+ \sum_{\rho=1\atop\rho\neq\mu}^{d-1} \f_{y,-\rho} \sum_{\nu=\rho+1\atop\nu\neq\mu}^d \f_{x,-\nu}\,.
\end{equation}
The Grassmann parity on the coarse grid is given by 
\begin{equation}
  \label{eq:FX}
  \f_{X,\nu}\equiv
  \begin{cases}
    \f_{x,\nu} & \text{for }\nu=-\mu\,,\\
    \f_{y,\nu} & \text{for }\nu=\mu\,,\\
    (\f_{x,\nu}+\f_{y,\nu}) \text{ mod }2 & \text{for }|\nu|\neq\mu\,.
  \end{cases}
\end{equation}
From \eqref{eq:sign} we see that we could assign individual sign factors to the fine-grid tensors $T_x$ and $T_y$, to the contraction, and to the coarse-grid tensor $\Tnew_X$. The numerical efficiency may depend on whether we apply the sign factors separately or as a whole.

\section{Proof of tensor decomposition after contraction}
\label{app:induction}

In this appendix we prove that \eqref{eq:Sum_split_of_tensor} is reproduced after one contraction step, before the OS-GHOTRG reduction and truncation are applied. This also yields the occupation numbers defined in \eqref{eq:coarse_occupations}.
We assume that \eqref{eq:Sum_split_of_tensor}
holds prior to a contraction step and
substitute it in the contraction \eqref{eq:just_contraction},
\begin{equation}
  (\Tnew_X)_{\fatj_X}
  = \sum_{\ns_x=0}^\nmax \sum_{\ns_y=0}^\nmax
	\sum_{j_{x,\mu}}\beta^{\ns_x+n_x+\ns_y+n_y}
	(T_x^{\ns_x})_{\j_x}
	(T_y^{\ns_y})_{\j_y} \sigma_{\f_x,\f_y}\,,
	\label{eq:contraction}
  \end{equation}
where we dropped the arguments of $n_x$ and $n_y$ for simplicity.  
Assuming that \eqref{eq:nx} holds for the current fine-grid tensors, we find 
\begin{align}
\ns_x+n_x+\ns_y+n_y
&= \ns_x+\ns_y  + \frac12 \sum_{\nu=\pm1}^{\pm d} (\nl_{x,\nu}+\nl_{y,\nu}) + \frac18 \sum_{\nu,\lambda=\pm1\atop|\nu|\ne|\lambda|}^{\pm d} (\nn{\lambda}_{x,\nu}+\nn{\lambda}_{y,\nu})
\notag\\
&= \underbrace{\ns_x + \ns_y + \frac12(\nl_{x,\mu}+\nl_{y,-\mu})}_{\ns_X} 
+  \frac12(\underbrace{\nl_{x,-\mu}}_{\nl_{X,-\mu}} + \underbrace{\nl_{y,\mu}}_{\nl_{X,\mu}}) 
+ \frac12\sum_{\nu=\pm1\atop|\nu|\ne\mu}^{\pm d}\Bigl[\underbrace{
\nl_{x,\nu}+\nl_{y,\nu} + \frac14
(\nn{\mu}_{x,\nu} + \nn{-\mu}_{y,\nu}  + \nn{\nu}_{x,\mu} + \nn{\nu}_{y,-\mu})}_{\nl_{X,\nu}}\Bigr]
\notag\\
&\quad 
 + \frac18 \sum_{\nu=\pm1\atop|\nu|\ne\mu}^{\pm d} (\underbrace{\nn{\nu}_{x,-\mu}}_{\nn{\nu}_{X,-\mu}} 
  + \underbrace{\nn{\nu}_{y,\mu}}_{\nn{\nu}_{X,\mu}} 
  + \underbrace{\nn{-\mu}_{x,\nu}}_{\nn{-\mu}_{X,\nu}} 
  + \underbrace{\nn{\mu}_{y,\nu}}_{\nn{\mu}_{X,\nu}}) 
  +\frac18\sum_{{{\nu,\lambda}=\pm1}\atop{{|\nu|,|\lambda|\ne\mu}\atop{|\nu|\ne|\lambda|}}}^{\pm d}(\underbrace{\nn{\lambda}_{x,\nu}+\nn{\lambda}_{y,\nu}}_{\nn{\lambda}_{X,\nu}}) \,,
  \label{eq:nX}
\end{align}
where the underbraces are identified with the site, link, and edge occupation numbers on the coarse grid.
Note that $(x,\mu)=(y,-\mu)$. 
When the hook conditions in \eqref{eq:pc} are satisfied, i.e., $\Delta_x\ne0$ and $\Delta_y\ne0$, we define the coarse-grid occupations as given in \cref{eq:coarse_occupations}, where the link occupation $\nl_{X,\nu}$ in the third line of \eqref{eq:nlXnu} was redefined using the hook conditions. 
We also use these definitions when the hook conditions are not satisfied and the tensor entries are zero.
Note that $\nl_{X,\nu}$ and $\nn{\lambda}_{X,\nu}$ are determined uniquely by $\fatj_{X,\nu}$. 
The exponent of $\beta$ in \eqref{eq:contraction} has the form $\ns_X+n_X$ if we define
\begin{equation}\label{eq:nxy}
  n_X(\fatj_X)\equiv 
  \frac12 \sum_{\nu=\pm1}^{\pm d} \nl_{X,\nu} + \frac18 \sum_{\nu,\lambda=\pm1\atop|\nu|\ne|\lambda|}^{\pm d} \nn{\lambda}_{X,\nu}\,,
\end{equation}
which is consistent with \eqref{eq:nx}.
Next, we split the sum over the contracted link index $j_{x,\mu}$ in \eqref{eq:contraction} into partial sums which run over the values of $j_{x,\mu}$ with the same link occupation $\nl_{x,\mu}(j_{x,\mu})$,
\begin{equation}\label{eq:splitsum}
	(\Tnew_X)_{\fatj_X}
	= \beta^{n_X} \sum_{\ns_x=0}^\nmax \sum_{\ns_y=0}^\nmax \sum_{m=0}^\nmax
	\beta^{\ns_X} 
    \sum_{{j_{x,\mu}}\atop{\nl_{x,\mu}=m}}
	(T_x^{\ns_x})_{\j_x}
	(T_y^{\ns_y})_{\j_y}
    \sigma_{\f_x,\f_y}
    \,,
\end{equation}
where we could pull out $\beta^{n_X}$ since $n_X$ does not depend on any of the summation variables and where the sum over $m$ terminates at $\nmax$ since this is the maximum link occupation after the reduction in the previous blocking step, see \eqref{eq:cutoff} in \cref{sec:reduction}.
We replace the sum over $m$ by a sum over  
\begin{equation}\label{app_nsX}
  \ns_X = \ns_x + \ns_y + m \,,
\end{equation}
see \eqref{eq:nsX2} with $\nl_{x,\mu}=m$, and then change the order of the sums
and adapt the ranges of the sums over $\ns_x$ and $\ns_y$ accordingly. Since we consider the strong-coupling expansion only up to $\beta^\nmax$, we truncate the resulting sum over $\ns_X$ at $\nmax$ to obtain
\begin{equation}\label{eq:recovered}
  	(\Tnew_X)_{\fatj_X}
	= \beta^{n_X}\sum_{\ns_X=0}^{\nmax} \beta^{\ns_X}
	\sum_{\ns_x=0}^{\ns_X}\sum_{\ns_y=0}^{\ns_X-\ns_x} \, 
	\sum_{j_{x,\mu}\atop\nl_{x,\mu}=\ns_X-\ns_x-\ns_y} 
	(T_x^{\ns_x})_{\j_x}
	(T_y^{\ns_y})_{\j_y}
    \sigma_{\f_x,\f_y}\,,
\end{equation}
which shows that
the tensor on the coarse site $X$ can be brought into the form \eqref{eq:Sum_split_of_tensor}, with $(\Tnew_X^{\ns_X})_{\fatj_X}$ given by \eqref{eq:MnsX}, the occupation numbers on the coarse grid defined in \eqref{eq:coarse_occupations}, and $n_X$ defined in \eqref{eq:nxy}.

Finally, it is straightforward to show that \eqref{eq:pcrg} is satisfied after the contraction step (by using $\Delta_x$, $\Delta_y$, and \eqref{eq:coarse_occupations}, or by using \cref{fig:coarsening} and invoking the hook conditions).

\section{Proof of link and site criteria}
\label{app:link-site}

After a contraction step, the partition function has the form of \eqref{eq:final_tensor_network} with $x$ replaced by $X$. Consider the product
\begin{equation}
  \prod_X (T_X^{\ns_X})_{\j_X}\ 
\end{equation}
for a single \vconfiguration $\bm{j}$. 
This configuration will contribute to the term of order $\beta^{N}$ in the partition function. Using \eqref{eq:Sum_split_of_tensor} and \eqref{eq:nx} we have
\begin{equation}\label{eq:N}
  N=\sum_X (\ns_X+n_X)  
  =\sum_X\bigg(\ns_X+\frac12\sum_{\mu=\pm1}^{\pm d}\nl_{X,\mu}+\frac18\sum_{{\mu,\nu=\pm1}\atop{|\mu|\ne|\nu|}}^{\pm d}\nn{\nu}_{X,\mu}\bigg)
  =\sum_X\bigg(\ns_X+\sum_{\mu=1}^{d}\nl_{X,\mu}+\frac12\sum_{{\mu,\nu=1}\atop{\mu\ne\nu}}^{d}\nn{\nu}_{X,\mu}\bigg)
\,,
\end{equation}
where in the last step we shifted some sites and used the link identity $(X,-\mu)=(X-\hat\mu,\mu)$ as well as the hook condition \eqref{eq:pc}.

The link criterion \eqref{eq:cutoff} is derived from \eqref{eq:N} for $\mu>0$ as follows,
\begin{align}
N&\ge\nl_{X,\mu}+\frac12\sum_{\nu=1\atop\nu\ne\mu}^d(\nn{\nu}_{X,\mu}+\nn{\mu}_{X,\nu})
+\frac12\sum_{\nu=1\atop\nu\ne\mu}^d(\nn{\nu}_{X-\hat\nu,\mu}+\nn{\mu}_{X-\hat\nu,\nu})\notag\\
&=\nl_{X,\mu}+\sum_{\nu=1\atop\nu\ne\mu}^d\nn{\nu}_{X,\mu}+\sum_{\nu=1\atop\nu\ne\mu}^d\nn{-\nu}_{X,\mu}
= \nl_{X,\mu}+\sum_{{\nu=\pm1}\atop{|\nu|\neq|\mu|}}^{\pm d}\nn{\nu}_{X,\mu} \,, \label{eq:link}
\end{align}
where in the first line we picked a subset of the terms in \eqref{eq:N}, and to obtain the second line we applied the link identity and the hook condition. Note that we derived \eqref{eq:link} for $\mu>0$, but using the link identity it is clear that it also holds for $\mu<0$. To have $N\le\nmax$, the link criterion \eqref{eq:cutoff} must be satisfied. Note that the link criterion is identical for all $T_X^{\ns_X}$.

The site criterion \eqref{eq:barnX} is derived from \eqref{eq:N} in a similar way,
\begin{align}
N&\ge \ns_X+\sum_{\mu=1}^d(\nl_{X,\mu}+\nl_{X-\hat\mu,\mu})
+\frac12\sum_{\mu,\nu=1\atop\mu\ne\nu}^d(\nn{\nu}_{X,\mu}+\nn{\nu}_{X-\hat\mu,\mu}+\nn{\nu}_{X-\hat\nu,\mu}+\nn{\nu}_{X-\hat\mu-\hat\nu,\mu})\notag\\
&= \ns_X+\sum_{\mu=1}^d(\nl_{X,\mu}+\nl_{X,-\mu})
+\frac12\sum_{\mu,\nu=1\atop\mu\ne\nu}^d(\nn{\nu}_{X,\mu}+\nn{\nu}_{X,-\mu}+\nn{-\nu}_{X,\mu}+\nn{-\nu}_{X,-\mu})\notag\\
&= \ns_X + \sum_{\mu=\pm1}^{\pm d} \nl_{X,\mu} 
+\sum_{{\mu,\nu=\pm1}\atop{|\mu|<|\nu|}}^{\pm d}\nn{\nu}_{X,\mu} \,.
\end{align}
The arguments for the first and the second line are the same as in the derivation of the link criterion.
In the last line we have also ordered $|\mu|<|\nu|$. To have $N\le\nmax$, the site criterion \eqref{eq:barnX} must be satisfied. Note that the site criterion is applied to each $T^{\ns_X}_X$ individually.

\section{Proof of tensor decomposition after tracing}\label{app:trace}

The manipulations in this section closely parallel those in \cref{app:induction}. Inserting \eqref{eq:Sum_split_of_tensor} in \eqref{eq:trace} yields
\begin{equation}\label{eq:TXd-1}
  (T_X)_{j_X}
  = \sum_{\ns_x=0}^{\nmax}\sum_{j_{x,\mu}}\beta^{\ns_x+n_x}(T_x^{\ns_x})_{j_x}\sigma_\mu\,,
\end{equation}
where here and below we have $j_{x,-\mu}=j_{x,\mu}$.
Using \eqref{eq:nx} and separating the occupation numbers related to the tracing direction $\mu$ we obtain  
\begin{equation}
  \ns_x+n_x=\underbrace{s_x+\frac12(\nl_{x,\mu}+\nl_{x,-\mu})}_{s_X}
  +\frac12\sum_{\nu=\pm1\atop{|\nu|\ne\mu}}^{\pm d}
  \Bigl[\underbrace{\nl_{x,\nu}+\frac14(\nn{\mu}_{x,\nu}+\nn{-\mu}_{x,\nu}+\nn{\nu}_{x,\mu}+\nn{\nu}_{x,-\mu})}_{\nl_{X,\nu}}\Bigr]
  +\frac18\sum_{{\nu,\lambda=\pm1}\atop{{|\nu|,|\lambda|\ne\mu}\atop{|\nu|\ne|\lambda|}}}^{\pm d}\underbrace{\nn{\lambda}_{x,\nu}}_{\nn{\lambda}_{X,\nu}}\,,
\end{equation}
where the underbraces are identified with the site, link, and edge occupation numbers on the $(d-1)$-dimensional lattice.
Using the link and hook conditions we define these occupations in Eq.~\eqref{eq:occupations_after_tracing}.
As before, we also use these definitions when the hook conditions are not satisfied and the tensor entries are zero. 
Note that $\nl_{X,\nu}$ and $\nn{\lambda}_{X,\nu}$ are determined uniquely by $j_{X,\nu}$.

The exponent of $\beta$ in \eqref{eq:TXd-1} has the form $s_X+n_X$ if we define $n_X$ as in \eqref{eq:nXd-1}.
Next, we split the sum over the link index $j_{x,\mu}$ in \eqref{eq:TXd-1} into partial sums over the values of $j_{x,\mu}$ with the same link occupation $\nl_{x,\mu}$,
\begin{equation}
  (T_X)_{j_X}
  = \beta^{n_X}\sum_{\ns_x=0}^{\nmax}\sum_{m=0}^{\nmax-\ns_x}\sum_{j_{x,\mu}\atop{\nl_{x,\mu}=m}}\beta^{\ns_X}  (T_x^{\ns_x})_{j_x}\sigma_{\mu}  \,,
\end{equation}
where $\beta^{n_X}$ could be pulled out since $n_X$ does not depend on any of the summation variables and where the upper limit on the sum over $m$ is reduced since we apply the site criterion \eqref{eq:barnX} in every blocking step.
We now eliminate $m$ in favor of $\ns_X=\ns_x+m$,
\begin{equation}
  (T_X)_{j_X}
  = \beta^{n_X}\sum_{\ns_x=0}^{\nmax}\sum_{\ns_X=\ns_x}^{\nmax}\sum_{j_{x,\mu}\atop{\nl_{x,\mu}=\ns_X-\ns_x}}\beta^{\ns_X}  (T_x^{\ns_x})_{j_x}\sigma_{\mu}
  = \beta^{n_X}\sum_{\ns_X=0}^{\nmax}\beta^{\ns_X}\sum_{\ns_x=0}^{\ns_X}\sum_{j_{x,\mu}\atop{\nl_{x,\mu}=\ns_X-\ns_x}} (T_x^{\ns_x})_{j_x}\sigma_{\mu}  \,,
\end{equation}
where we switched the order of the sums over $\ns_x$ and $\ns_X$ in the last step.
This shows that $(T_X)_{j_X}$ can be written in the form of \eqref{eq:Sum_split_of_tensor} with $(T_X^{\ns_X})_{j_X}$ given in \eqref{eq:TXsXd-1}.
Therefore the tensor decomposition is preserved after the tracing.

\section{Validating the coefficients of \cref{sec:matching}}
\label{app:Trans_inv}

Since $\nR \geq \lfloor \nmax/2 \rfloor$, all partitions that appear in $Z$ also appear in $\X$. We now show that the coefficients of terms in $Z_n$ match those of $a_n \X_n + b_n \Y_n$ if $a_n$, $b_n$, and $c_n$ are given by \eqref{eq:abc_k}.

Consider a specific partition
$(n) = (n_1, \ldots, n_1, n_2, \ldots n_2, \ldots n_m, \ldots, n_m)$
of $n$ as defined in \cref{sec:contributions}, where $k_i$ is the multiplicity of $n_i$ and $n= \sum_{i=1}^m k_i n_i$. 
The contribution of this partition to $Z_n$ is $Z_{(n)}$.
To compute the contribution of this partition to $a_n \X_n + b_n \Y_n$ we distinguish two cases.

The first case is $n_1 \leq \nR$. Since the $n_i$ are ordered, this implies $n_i\leq \nR$ for all $i$. Then all terms contained in $\X_n$ that are related to $(n)$ by permutation of the $n_i$ are given by
\begin{equation}
  \sum_{i=1}^m c_{n_i} Z_{(n_i|\, n \backslash n_i)} \stackrel{\eqref{eq:Zn}}{=} \frac{Z_{(n)}}{V} \sum_{i=1}^m c_{n_i} k_i
  \stackrel{\eqref{eq:abc_k}}{=} \frac{Z_{(n)}}{V} \sum_{i=1}^m \frac{n_i k_i}{\nmax}
  =  \frac{n}{V \nmax} Z_{(n)}\,.
\end{equation}
Since $n_i\leq \nR$ for all $i$, the term in $\Y_n$ related to $(n)$ is given by $Z_{(n)}$. Therefore, in $a_n \X_n + b_n \Y_n$, all terms related to $(n)$ are given by
\begin{equation}
  a_n  \frac{n}{V \nmax} Z_{(n)} + b_n Z_{(n)} \stackrel{\eqref{eq:abc_k}}{=} Z_{(n)}\,.
\end{equation}

The second case is $n_1 > \nR$. Due to $\nR \geq \lfloor \nmax/2 \rfloor$, this  implies $k_1=1$ and $n_i \leq \nR$ for $i\geq 2$.
Then the only term in $\X_n$ related to $(n)$ is given by
\begin{equation}
  c_{n_1} Z_{(n_1|\,n \backslash n_1)} \stackrel{\eqref{eq:abc_k}}{=} Z_{(n_1|\,n \backslash n_1)} \stackrel{\eqref{eq:Zn}\text{ for }k_1=1}{=} \frac1V Z_{(n)}\,.
\end{equation}
Any tuple obtained from $(n)$ by a nontrivial permutation cannot appear in $\X_n$, since $n_1>\nR$. For the same reason, $\Y_n$ does not contain terms related to $(n)$.
Therefore, in $a_n \X_n + b_n \Y_n$ with $n\geq n_1 > \nR$, all terms related to $(n)$ are given by
\begin{equation}
  \frac{a_n} V Z_{(n)} \stackrel{\eqref{eq:abc_k}\text{ for }n>\nR}{=} Z_{(n)}\,.
\end{equation}

In both cases, the sum of all terms in $a_n X_n + b_n Y_n$ that are related to $(n)$ is given by $Z_{(n)}$. Since the arguments above apply for any partition $(n)$ of $n$, this concludes the proof.

\section{Asymptotic results for large chemical potential}
\label{app:large_mu}

We first consider the case of $\Nf=1$ and discuss the generalization to arbitrary $\Nf$ at the end of this section.
The strong-coupling expansion of the partition function for $\Nf=1$ has the form
\begin{equation}
  Z(\mu,m,\beta)=\sum_{n=0}^{\infty}Z_n(\mu,m)\beta^n\,.
\end{equation}
Note, however, that the partition function can also be written as
\begin{equation}\label{eq:Z_cosh_Nf1}
  Z(\mu,m,\beta)=\sum_{k=0}^{\Vs}p_{k}(m,\beta)\cosh(k\Nc \Lt \mu)\,,
\end{equation}
where $\Lt$ is the temporal extent of the $d$-dimensional space-time lattice and $\Vs$ is the spatial volume. To explain \eqref{eq:Z_cosh_Nf1}, we first note that at infinite coupling, the $\mu$ dependence of $Z$ comes from closed baryon and antibaryon loops winding around the torus in the time direction \cite{Rossi:1984cv,Karsch:1988zx}. Every winding contributes a factor of $e^{\pm\Nc \Lt\mu}$ to a configuration. Since baryon loops are self-avoiding, the maximum winding number is $\Vs$.
For every configuration with net winding number $k$ there is a corresponding configuration with baryon and antibaryon loops interchanged and winding number $-k$, and therefore we obtain $\cosh(k\Nc\Lt\mu)$.
For nonzero $\beta$ the baryon loops are locally disturbed by plaquettes (more precisely, one or more quark lines in the baryon go around the plaquettes), but the functional form of the $\mu$ dependence remains the same. Analytical results for a $2\times2$ lattice are given in  \cite[App.~B]{Samberger:2025hsr}.

We now perform a strong-coupling expansion of the coefficients,
\begin{equation}
  p_k(m,\beta)=\sum_{n=0}^{\infty}p_{k,n}(m)\beta^n\,,
\end{equation}
and obtain
\begin{equation}
  Z_n(\mu,m)=\sum_{k=0}^{\Vs}p_{k,n}(m)\cosh(k\Nc \Lt \mu)\,.
\end{equation}
In the limit $|\mu|\rightarrow\infty$ the sum is dominated by the $k=\Vs$ term, for which we show in \cref{app:large_mu_pure_gauge} that
\begin{equation}\label{eq:pVs}
  p_{\Vs}(m,\beta)=2Z^{\text{PG}}(\beta)
\end{equation}
independent of $m$, where $Z^{\text{PG}}$ is the partition function of the pure gauge theory (with the same lattice size and number of colors), whose strong-coupling expansion reads 
\begin{equation}
	Z^{\text{PG}}(\beta)=\sum_{n=0}^{\infty}Z_n^{\text{PG}}\beta^n\,.
\end{equation}
This leads to
\begin{equation}\label{eq:pVsn=2PG}
  p_{\Vs,n}(m)=p_{\Vs,n}=2Z^\text{PG}_n\,,
\end{equation}
and with $V=\Lt\Vs$ we obtain in the $|\mu|\rightarrow\infty$ limit 
\begin{equation}\label{eq:Zn=ZnPG}
  Z_n(\mu,m)\rightarrow 2Z^\text{PG}_n\cosh(\Nc V \mu)\rightarrow Z^\text{PG}_n\exp(\Nc V |\mu|)\,,
\end{equation}
provided that $p_{\Vs,n}\neq0$. This yields
\begin{equation}\label{eq:ratio}
  \frac{Z_n(\mu,m)}{Z_0(\mu,m)}\rightarrow \frac{Z^\text{PG}_n}{Z^\text{PG}_0}
\end{equation}
for any $n\geq0$.\footnote{If $Z_n^\text{PG}=0$, which happens in particular for $n=1$, we obtain $Z_n(\mu,m)\rightarrow p_{k,n}(m)\cosh(k\Nc \Lt  \mu)$, where $k<V_s$ is the largest value of $k$ for which $p_{k,n}(m)\ne0$. In this case $\cosh(k\Nc \Lt \mu)/\cosh(\Vs \Nc \Lt \mu)\rightarrow0=Z_n^\text{PG}$ for $k<\Vs$, and hence the relation \eqref{eq:ratio} still applies.} 
The ratio is well-defined since $Z^\text{PG}_0=\text{vol}(\SU(\Nc))^{Vd}=1$.
Note that for $n\ge1$ the coefficients $f_n$ are polynomials in the $\Zbar_n=Z_n/Z_0$, see \eqref{eq:logZexp} and \eqref{logZexp}. We thus obtain
\begin{equation}\label{eq:fntoPG}
  f_n(\mu,m)\rightarrow f_n^{\text{PG}}+\Nc |\mu|\delta_{n,0}\quad\text{for $|\mu|\rightarrow\infty$}\,,
\end{equation}
where the Kronecker delta takes into account the different definition of $f_0$.
From \eqref{eq:fntoPG} we can obtain the behavior of the chiral condensate and the quark-number density in this limit. Since there is no dependence on $m$, all coefficients of the chiral condensate approach zero, i.e.,\footnote{Strictly speaking, the limit $|\mu|\to\infty$ and the derivative with respect to $m$ may not commute. However, a more careful analysis in which we first take the derivative and then the limit confirms \eqref{eq:analytic_limit_sigma_mu=infty}.} 
\begin{equation}\label{eq:analytic_limit_sigma_mu=infty}
	\Sigma_n(\mu,m)\rightarrow0\quad\text{for $|\mu|\rightarrow\infty$}\,.
\end{equation}
For the quark-number density we obtain
\begin{equation}\label{eq:analytic_limit_rhoi_mu=infty}
  \rho_n(\mu,m)\rightarrow\pm\Nc\delta_{n,0}\quad
  \text{for $\mu\rightarrow\pm\infty$}\,.
\end{equation}

For arbitrary $\Nf$, \eqref{eq:Z_cosh_Nf1} generalizes to
\begin{equation}
  Z(\vec\mu,\vec m,\beta)=\sum_{k_1,\ldots,k_{\Nf}=-\Nc\Vs}^{\Nc\Vs}p_{\vec k}(\vec m,\beta)\exp\bigl(\Lt\,\vec k\!\cdot\!\vec\mu\bigr)\,,  
\end{equation}
where $\vec k=(k_1,\ldots,k_{\Nf})$, $\vec\mu=(\mu_1,\ldots,\mu_{\Nf})$, and $\vec m=(m_1,\ldots,m_{\Nf})$. The $p_{\vec k}$ are defined in a slightly different way compared to \eqref{eq:Z_cosh_Nf1} and are only nonzero if $\sum_{f=1}^{\Nf}k_f=0\mod\Nc$.  In the limit $|\mu_f|\to\infty$ for one flavor $f$, the arguments proceed in analogy to the one-flavor case, except that instead of $Z_n^\text{PG}$ we obtain the coefficients $Z_n^{(\Nf-1)}$ of the partition function for the remaining $\Nf-1$ flavors in \eqref{eq:Zn=ZnPG}, see \cite{Samberger2026}. Therefore \eqref{eq:analytic_limit_sigma_mu=infty} and \eqref{eq:analytic_limit_rhoi_mu=infty} remain valid for flavor $f$.

\section{\boldmath The function $p_{\Vs}(m,\beta)$}\label{app:large_mu_pure_gauge}

The purpose of this appendix is to prove the relation \eqref{eq:pVs}.
We can read off the expression for $p_{\Vs}$ from \cite[Eqs.~(2.1) and (2.5)]{Samberger:2025hsr} for $\Nf=1$ by identifying the coefficient of $\exp(\Nc V \mu)$ and obtain
\begin{equation}
  \label{eq:barpVs_def}
  \frac12 p_{\Vs}(m,\beta) = \int \left[\prod_{x,\mu} dU_{x,\mu}\right]
  \left[\prod_x\mathcal{B}_x\right]
  e^{\sum\limits_{x,\mu\neq\nu} \SG_{x,\mu\nu}}
\end{equation}
with\footnote{The Grassmann differentials have been reordered compared to \cite{Samberger:2025hsr}, but this does not result in a sign change since the lattice extent is even in every direction.}
\begin{align}\label{eq:B=1}
  \mathcal{B}_x&\equiv\frac{1}{\Nc!}\int\left[\coprod\limits_{i=1}^{\Nc}d\psi_{x+\hat1}^{i} \right]
  \left[\prod_{i=1}^{\Nc}d\bar\psi_{x}^{i} \right]
  \left(\bar\psi_x U_{x,1} \psi_{x+\hat1}\right)^{N_{\text{c}}} \notag\\
  &=
  \frac{1}{\Nc!}\int\left[\coprod\limits_{i=1}^{\Nc}d\psi_{x+\hat1}^{i} \right]
  \left[\prod_{i=1}^{\Nc}d\bar\psi_{x}^{i} \right]
  \left[\coprod_{k=1}^{\Nc}
	\bar\psi^{i^k_{x,1}}_x\right]
  \left[\prod_{k=1}^{\Nc}
	U^{i^k_{x,1}j^k_{x,1}}_{x,1} \right]
  \left[\prod_{k=1}^{\Nc}
	\psi^{j^k_{x,1}}_{x+\hat1}\right]
  \notag\\
  &=
  \frac{1}{\Nc!}\varepsilon_{i^1_{x,1}\dotsb i^{\Nc}_{x,1}}			
  \left[\prod_{k=1}^{\Nc}
	U^{i^k_{x,1}j^k_{x,1}}_{x,1} \right]
  \varepsilon_{j^1_{x,1}\dotsb j^{\Nc}_{x,1}}
  =
  \frac{1}{\Nc!}\det(U_{x,1})\,\varepsilon_{j^1_{x,1}\dotsb j^{\Nc}_{x,1}}
  \varepsilon_{j^1_{x,1}\dotsb j^{\Nc}_{x,1}}	=1	\,,
\end{align}
where sums over repeated color indices are implied. We integrated out the Grassmann variables using \cite[Eq.~(A.23)]{Samberger:2025hsr} and used the Leibniz formula for the determinant in combination with permutations of the columns. In the last step we used $\det(U_{x,1})=1$ and contracted the Levi-Civita tensors, which results in a factor of $\Nc!$.

\Cref{eq:B=1} shows that the $\Vs$ parallel baryon loops in the time direction result in a constant background, independent of the gauge field. Therefore \eqref{eq:barpVs_def} is simply the partition function of the pure gauge theory with the same lattice size and number of colors as the original theory. This concludes the proof.

\section{\boldmath Asymptotic results for large mass}
\label{app:large_mass}

For $\Nf=1$ and large $m$, the partition function can be approximated, up to $\mathcal O(\beta^{\Lt})$ corrections, as
\begin{equation}\label{eq:Zlarge_mass}
  Z_\text{factorized}(\mu,m,\beta)=\left(\left(2m\right)^{\Nc \Lt}+2\cosh(\Nc \Lt\mu)\right)^{\Vs}Z^{\text{PG}}(\beta)\,.
\end{equation}
To derive this result, we start from \eqref{eq:Z_cosh_Nf1} and determine, for given $k$, the leading-order term in $p_k(m, \beta)$ for large $m$. For powers of $\beta$ smaller than $\Lt$, this term is obtained by configurations which have $k$ baryon loops (or $k$ antibaryon loops) with all links in the time direction and winding around the torus once, and all other lattice sites occupied by mass terms. For all other configurations, e.g., with mesonic contributions, or baryon lines in a spatial direction, or combinations of baryon and antibaryon windings, the number of lattice sites that can be occupied by mass terms is reduced. Therefore such configurations do not contribute to $p_k$ at leading order in $m$.
For a configuration which has only mass terms and $k$ baryon (or antibaryon) loops, the gauge field only occurs in contributions from the gauge action, see \eqref{eq:B=1}. 
For powers of $\beta$ larger than or equal to $\Lt$, there are also contributions from other configurations,\footnote{An example for such a configuration, for $\Nc=3$ and an $\Lt\times2$ lattice, would be two quark loops in the time direction with spatial coordinate~$1$, one quark loop in the time direction with spatial coordinate $2$, and plaquette occupation $1$ for all plaquettes $(x,12)$ starting at sites $x=(t,1)$ with $t=1,\ldots,\Lt$.} but we do not include them explicitly since in this paper we are only interested in an expansion of $Z$ up to $\beta^{\nmax}$, and typically $\nmax<\Lt$.
We therefore have for large $m$
\begin{equation}\label{eq:pk}
  p_k(m, \beta) \to (2-\delta_{k,0}) \binom{\Vs}k   (2m)^{\Nc \Lt (\Vs-k)} Z^\text{PG}(\beta)+\mathcal O(\beta^{\Lt})
\end{equation}
since there are $\binom{\Vs}k$ possibilities to distribute the $k$ parallel baryon loops on the lattice. The factor of $2-\delta_{k,0}$ comes from $e^{k \Nc \Lt \mu} +e^{- k \Nc \Lt \mu} = 2  \cosh(k \Nc \Lt \mu)$, which sums configurations with $k$ baryon loops and configurations with $k$ antibaryon loops. 
This results in
\begin{equation}\label{eq:Zasymp}
  Z(\mu,m,\beta) \to  Z_\text{asymptotic}(\mu,m,\beta) \equiv Z^\text{PG}(\beta) \sum_{k=0}^{\Vs} \binom{\Vs}k (2m)^{\Nc \Lt (\Vs-k)} (2-\delta_{k,0}) \cosh(k \Nc \Lt \mu)  
\end{equation}
up to $\mathcal O(\beta^{\Lt})$ corrections.
Let us now show that the large-$m$ behavior of \eqref{eq:Zasymp} is reproduced by \eqref{eq:Zlarge_mass}. Expanding \eqref{eq:Zlarge_mass} gives
\begin{equation}
  Z_\text{factorized}(\mu,m,\beta) = Z^\text{PG}(\beta) \sum_{k=0}^{\Vs} \binom{\Vs}k (2m)^{\Nc \Lt (\Vs-k)} 2^k\cosh^k( \Nc \Lt \mu)\,.   
\end{equation}
Defining $x=\Nc \Lt \mu$ and using 
\begin{equation}\label{eq:cosh^k}
  \cosh^k(x) = \frac1{2^k}\left(e^x+e^{-x}\right)^k = \frac1{2^k} \sum_{j=0}^k \binom k j e^{(k-2j) x}
   = \frac1{2^k} \sum_{j=0}^k \binom k j \cosh\bigl((k-2j) x\bigr)\,,
\end{equation}
we can now organize the terms in $Z_\text{factorized}$ in complete analogy to \eqref{eq:Z_cosh_Nf1}. Next, we approximate $Z_\text{factorized}$ for large $m$. For the coefficient of $\cosh(k \Nc \Lt \mu)$ in $Z_\text{factorized}$, we again want to keep only the leading-order term for large $m$. This term is obtained by setting $j=0$ or $j=k$ in \eqref{eq:cosh^k}.
Hence, to obtain the large-$m$ behavior of $Z_\text{factorized}$, we can replace $2^k\cosh^k(x) \to (2-\delta_{k,0})\cosh(k x)$. This shows that $Z_\text{factorized} \to Z_\text{asymptotic}$ for large $m$.

For the chiral condensate $\Sigma=(\partial_m\ln Z)/V$ in the large-$m$ limit and up to $\mathcal O(\beta^{\Lt})$ corrections we obtain from \eqref{eq:Zlarge_mass} 
\begin{equation}\label{eq:Sigma_largem}
  \Sigma(\mu,m)=\frac{2\Nc(2m)^{\Nc \Lt-1}}{(2m)^{\Nc \Lt}+2\cosh(\Nc \Lt\mu)}
  =\bar b\frac{2\sqrt{\left(\frac12(2m)^{\Nc \Lt}\right)^2-1}}{\frac12(2m)^{\Nc \Lt}+\cosh(\Nc \Lt\mu)}
  =\bar b\bigl(\tanh(\bar a(\mu+\bar\mu^\text{c}))-\tanh(\bar a(\mu-\bar\mu^\text{c}))\bigr)\,,
\end{equation}
where we defined
\begin{equation}
  \bar b\equiv\frac{\Nc(2m)^{\Nc \Lt-1}}{2\sqrt{\left(\frac12(2m)^{\Nc \Lt}\right)^2-1}}\rightarrow\frac{\Nc}{2m}\,,\quad
  \bar a\equiv\frac{\Nc \Lt}{2}\,, \quad
  \bar\mu^\text{c}\equiv\frac{1}{\Nc \Lt}\text{arccosh}\left(\frac12\left(2m\right)^{\Nc \Lt}\right)\rightarrow\ln (2m)\,,
\end{equation}
and used the general formula
\begin{equation}\label{eq:tanh_general_formula}
  \frac{2\sinh(2x)}{\cosh(2x)+\cosh(2z)}=\tanh(x+z)+\tanh(x-z)\,.
\end{equation}
Note that \eqref{eq:Sigma_largem} has the structure of the model function given in \cref{footnote:cc_model}.

Similarly, for the quark-number density $\rho=(\partial_\mu\ln Z)/V$ we obtain in the large-$m$ limit and up to $\mathcal O(\beta^{\Lt})$ corrections 
\begin{equation}\label{eq:rho_largem}
  \rho(\mu,m)=\frac{\Nc}{2}\frac{2\sinh(\Nc \Lt\mu)}{\frac12(2m)^{\Nc \Lt}+\cosh(\Nc \Lt\mu)}
  =\frac{\Nc}2\bigl(\tanh(\bar a(\mu+\bar\mu^{\text{c}}))+\tanh(\bar a(\mu-\bar\mu^{\text{c}}))\bigr)\,,
\end{equation}
where we again used \eqref{eq:tanh_general_formula}. 
Note that critical chemical potential and transition sharpness are identical for both observables.
\Cref{eq:rho_largem} has the structure of the model function given in \cref{footnote:rho_model}.

For $m\to\infty$, we obtain $\Sigma\to0$ from \eqref{eq:Sigma_largem} since $\bar b\to0$ in this limit, and we obtain $\rho\to0$ from \eqref{eq:rho_largem} at fixed $\mu$ since $\muc\to\infty$. This implies that the coefficients $\Sigma_n$ and $\rho_n$ go to zero for $n<\Lt$.

\section{Coefficients of the strong-coupling expansion for the observables from tanh models}\label{app:Coeff_exp}

For both model functions \eqref{ccfit} and \eqref{eq:fit-ansatz} we first expand the argument of the tanh in $\beta$,
\begin{equation}
  a(\beta) (\mu - \mu^\text{c}(\beta)) = \sum_{n=0}^{\nmax} \bar \mu_n \beta^n + \mathcal O\left(\beta^{\nmax +1}\right)
\end{equation}
with 
\begin{subequations}\label{eq:}
  \begin{align}
    \bar \mu_0 &  = a_0 (\mu - \mu^\text{c}_0) \,,\\
    \bar \mu_1 &  = a_1 (\mu -\mu^\text{c}_0)-a_0 \mu^\text{c}_1 \,, \label{eq:bar_mu1}\\
    \bar \mu_2 &  = a_2 (\mu -\mu^\text{c}_0)-a_1 \mu^\text{c}_1-a_0 \mu^\text{c}_2 \,,\\
    \bar \mu_3 &  = a_3 (\mu -\mu^\text{c}_0)-a_2 \mu^\text{c}_1-a_1 \mu^\text{c}_2-a_0 \mu^\text{c}_3\,.
  \end{align}
\end{subequations}
 Here and below we omit the labels $\Sigma$ and $\rho$ for the parameters $a_n$ and $\mu^\text{c}_n$. 
 Using $t\equiv \tanh(\bar \mu_0)$ and $s\equiv \sech(\bar \mu_0)$ we obtain
 \begin{subequations}\label{eq:sigma_n}
   \begin{align}
     \Sigma_0^{\text{model}}(\mu) &= b_0(1-t) \,,\\
     \Sigma_1^{\text{model}}(\mu) &= b_1(1-t)- b_0 \bar \mu_1  s^2 \,,\\
     \Sigma_2^{\text{model}}(\mu) &= b_2(1-t) -b_1 \bar\mu_1 s^2 - b_0 \left(\bar\mu_2 -\bar \mu_1^2 t\right) s^2 \,,\\
     \Sigma_3^{\text{model}}(\mu) &= b_3(1-t)-b_2 \bar\mu_1 s^2 - b_1 \left(\bar\mu_2 -\bar \mu_1^2 t\right) s^2 - b_0 \left(\bar\mu_3 -2 \bar\mu_2 \bar\mu_1 t -\bar\mu_1^3 \left(s^2-\frac 23\right) \right)s^2 \,.
  \end{align}
 \end{subequations}
 The results for $\rho_n$ can be obtained by replacing $b_0 \to \frac32$, $b_n \to 0$ for $n\geq 1$, and $\bar \mu_n \to - \bar \mu_n$ for all $n$ (resulting also in $t\to-t$) in the expressions for $\Sigma_n$. This yields
 \begin{subequations}\label{eq:rho_n}
   \begin{align}
     \rho_0^{\text{model}}(\mu) &= \frac32(1+t) \,,\\
     \rho_1^{\text{model}}(\mu) &= \frac32 \bar \mu_1  s^2 \,,\label{eq:rho_1}\\
     \rho_2^{\text{model}}(\mu) &= \frac32 \left(\bar\mu_2 -\bar \mu_1^2 t\right) s^2 \,,\\
     \rho_3^{\text{model}}(\mu) &= \frac32 \left(\bar\mu_3 -2 \bar\mu_2 \bar\mu_1 t -\bar\mu_1^3 \left(s^2-\frac 23\right) \right)s^2 \,.
  \end{align}
 \end{subequations}

Using these results we can explain the low sensitivity of the model coefficients in \eqref{eq:sigma_n} and \eqref{eq:rho_n} on $a_n/a_0$ for $n\ge1$ and on $\muc_n$ for $n\ge2$ when $a_0$ becomes large, which is the case for large volume. 
For simplicity we focus on $\rho$, but similar arguments can be made for $\Sigma$.
Let us first consider $\rho_1$, suppressing the superscript ``model'' for brevity. Because of $\sech(\bar \mu_0)$, $\rho_1$ becomes exponentially suppressed when $|\bar \mu_0|\gg 1$. We therefore only need to consider $|\bar \mu_0| \le \mathcal O(1)$, i.e., the vicinity of the phase transition, in \eqref{eq:bar_mu1}, which can be written as 
\begin{equation}
  \bar \mu_1  = \frac{a_1}{a_0} \bar \mu_0 -a_0 \mu^\text{c}_1\,.
\end{equation}
We see that the first term becomes less and less relevant when $a_0$ becomes large at fixed $a_1/a_0$, provided that $\mu^\text{c}_1 \neq 0$. Therefore $\rho_1$ in \eqref{eq:rho_1} becomes increasingly insensitive to $a_1/a_0$ in this scenario.

Similarly, for large $a_0$ we see that $\rho_2$  is dominated by the $(a_0 \muc_1)^2$ term originating from $\bar \mu_1^2$ and that $\rho_3$ is dominated by the $(a_0\muc_1)^3$ term originating from $\bar \mu_1^3$. This pattern continues to larger values of $n$, and therefore all $\rho_n$ for $n\geq 1$ are dominated by $a_0^n$ and insensitive to $a_1/a_0,\ldots,a_n/a_0$ when $a_0$ becomes large for large $V$.

The same argument also applies to $\muc_n$ for $n\ge2$. For example, the subleading terms in $\rho_2/s^2$ are $a_1 \muc_1$ and $a_0\muc_2$. Both terms are suppressed by $1/a_0$ compared to the leading term $(a_0\muc_1)^2$. Hence determining $\muc_2$ or $a_1$ from $\rho_2$ should be equally hard. A more careful analysis of the zeros of $\rho_2$ (which we do not present here) does not change this expectation.
To conclude, in a combined fit for all $\rho_n$, it appears that $a_n/a_0$ and $\muc_{n+1}$  should be equally difficult to determine for large $V$, provided that all $a_n/a_0$ and $\muc_n$ have finite large-$V$ limits which are of similar magnitude.

\section{\boldmath Interpolation between small and large $\mu$ for large $\Lt$}
\label{app:interpolation}

The purpose of this appendix is to derive interpolating formulas that turn out to describe the data very well in many cases, with limitations discussed below.     
From \eqref{eq:Z_cosh_Nf1} and \eqref{eq:pVs} we see that for $|\mu| \to \infty$ we have $Z= 2 Z^\text{PG}(\beta) \cosh(\Nc V \mu) (1+ \mathcal O(e^{- \Nc \Lt |\mu|}))$.
On the other hand, in the vicinity of $\mu=0$ and for large $\Lt$ we expect $Z$ to remain approximately constant, i.e., $Z(\mu, m, \beta) \approx Z(\mu=0, m, \beta) \approx p_0(m,\beta)$ for the following reason. The coefficients $p_k$ in $\eqref{eq:Z_cosh_Nf1}$ have to decrease exponentially with $k$ since $p_k$ contains powers of $m$ and combinatorial factors for distributing mass terms and mesons on up to $(V_\text{s} -k) \Lt$ lattice sites, i.e., $\ln p_k \sim (V_\text{s} -k) \Lt$. Hence, $Z(\mu)$ will deviate from its value at $\mu=0$ only when $|\mu|$ is sufficiently large for $\cosh(k \Nc \Lt \mu)$ to compensate the exponential suppression of $p_k$. We expect this to happen in the vicinity of the phase transition.
 
A simple interpolation between the two regimes is provided by
\begin{equation}\label{eq:ZIP}
  Z_\text{IP}(\mu,m,\beta) = c(m,\beta) + 2 Z^\text{PG}(\beta) \cosh(\Nc V \mu) \quad \text{with}\quad  c(m,\beta) = Z(\mu=0, m, \beta) - 2 Z^\text{PG}(\beta)\,.
\end{equation}
For the chiral condensate and the quark-number density this yields the symmetrized (see \cref{footnote:cc_model}) and antisymmetrized (see \cref{footnote:rho_model}) tanh model, respectively,  
\begin{align}
  \Sigma_\text{IP}(\mu,m,\beta)
  &=b_{\text{IP}}\left(\tanh(a_{\text{IP}}(\mu+\muc_{\text{IP}}))-\tanh(a_{\text{IP}}(\mu-\muc_{\text{IP}}))\right),\\
  \rho_\text{IP}(\mu,m,\beta) &= \frac{\Nc}2 \left(\tanh(a_\text{IP}(\mu+\muc_\text{IP}))+\tanh(a_\text{IP}(\mu-\muc_\text{IP}))\right),
\end{align}
where we used \eqref{eq:tanh_general_formula} and defined
\begin{align}
  b_{\text{IP}}&\equiv\frac{\partial_m c(m,\beta)}{4V Z^\text{PG}(\beta)\sqrt{\left(\frac{c(m,\beta)}{2Z^\text{PG}(\beta)}\right)^2-1}}
  \approx\frac{\partial_mc(m,\beta)}{2Vc(m,\beta)}\approx\frac12\Sigma(\mu=0,m,\beta) \,,\\ 
  a_{\text{IP}}&\equiv\frac{\Nc V}{2}\,, \label{eq:a_ip}\\
  \muc_{\text{IP}}&\equiv\frac{1}{\Nc V}\text{arccosh}\frac{c(m,\beta)}{2Z^\text{PG}(\beta)}
  \approx \frac1{\Nc V} \ln \frac{c(m, \beta)}{Z^\text{PG}(\beta)}\,.
\end{align}
For the approximations we used $c\gg2Z^\text{PG}$, which follows from $Z(\mu=0)\approx p_0\gg p_{\Vs}=2Z^\text{PG}$. 
We compute $Z(\mu=0,m,\beta)$ numerically using OS-GHOTRG and use $Z^\text{PG}(\beta) = 1+ \frac12d(d-1)V(1+\delta_{\Nc,2})(\beta/2\Nc)^2+\mathcal O(\beta^3)$, which is obtained from the definition of the gauge action \cite{Samberger:2025hsr} using the results of \cite{Creutz:1978ub}.
As usual, we expand $b_\text{IP}$ and $\muc_\text{IP}$ in $\beta$, with coefficients $b_{\text{IP},n}$ and $\muc_{\text{IP},n}$, respectively.

\begin{figure}
  \centering{
    \hfill
    \includegraphics[width=.4\textwidth]{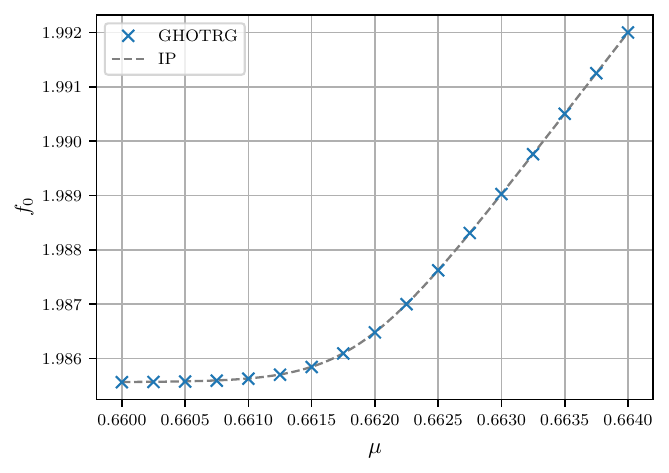}
    \hfill
    \includegraphics[width=.4\textwidth]{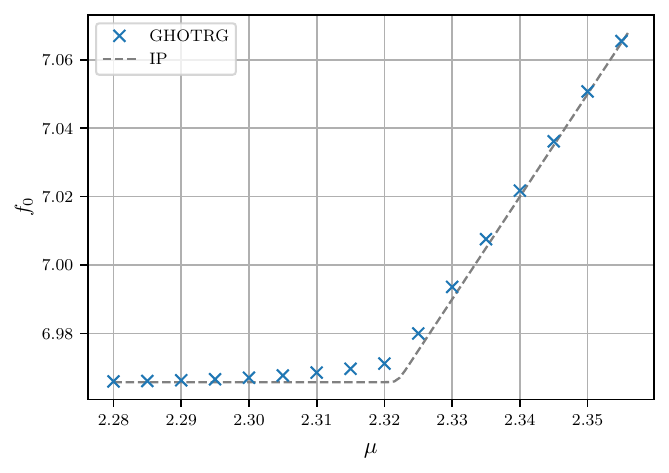}
    \hfill
  }
  \caption{Numerical results for $f_0=(\ln Z_0)/V$ obtained from GHOTRG on a $32\times32$ lattice with bond dimension $D=60$ and interpolating formula \eqref{eq:ZIP} for $m=0.5$ (left) and $m=5$ (right).}
  \label{fig:f0_ip}
\end{figure}

In \cref{fig:f0_ip} we compare numerical results for $Z$ obtained in the infinite-coupling limit from the GHOTRG method with the interpolating formula \eqref{eq:ZIP}. 
We see that the interpolation works very well away from the phase transition, but an 
interesting question is whether it also provides a reasonable approximation to $Z$ in the vicinity of the transition. In \cref{fig:f0_ip} we see that the answer to this question depends on the parameters of the system such as the mass.
We can distinguish two scenarios for the transition:
\begin{enumerate}[label=(\alph*)]
\item For any $\mu<\muc$, the $k=0$ term is the dominating term in \eqref{eq:Z_cosh_Nf1}, while for any $\mu>\muc$ the $k=\Vs$ term dominates. In this case we expect the interpolation to provide a reasonable approximation to the true $Z$ also across the phase transition.
\item In the vicinity of $\muc$, terms other than $k=0$ or $k=\Vs$ make the largest contribution to \eqref{eq:Z_cosh_Nf1}. 
In this case, keeping only the $k=0$ or $k=V_\text{s}$ term will not provide a reasonable approximation to $Z$ when $\mu\approx \muc$. We then expect the phase transition to be less sharp with $a<a_\text{IP}$.
 \end{enumerate}
Which scenario occurs in practice depends on $m$, $L$, and $\beta$ through the precise values of $p_k(m, \beta)$.
For large $m$, we have scenario (b) since  in this case $e^{\muc} \approx 2m$ (see \cref{app:large_mass}) and $p_k \sim (2m)^{\Nc \Lt (\Vs-k)}$, which implies that in \eqref{eq:Z_cosh_Nf1} all terms are of the same order of magnitude. (At $\mu=\muc$, the dominating terms are those with $k\approx \Vs/2$ because of the binomial coefficient in \eqref{eq:Zasymp}.) This is reflected in $a=\Nc \Lt/2$ for large $m$ (see \cref{app:large_mass}). Hence, the actual transition is less sharp than that of $\Sigma_\text{IP}$ and $\rho_\text{IP}$, where $a_\text{IP}=\Nc V/2$. This explains why the interpolation deviates from the data for the larger mass in \cref{fig:f0_ip} in the vicinity of the phase transition.
Nevertheless, we see in \cref{fig:coeff_vs_m} (left) that $\muc_\text{IP}$ still provides an excellent approximation to $\muc$ for all $m$ (for lattice size $16 \times 16$). Note that for large $m$ we have $c(m,\beta)\to (2m)^{\Nc V}$, resulting in $\muc_\text{IP} \to \ln (2m)$, which is consistent with $\muc$ of \cref{app:large_mass}.

\bibliographystyle{elsarticle-num.bst}
\bibliography{qcd_os}

\end{document}